\documentclass[12pt,english,review]{elsarticle}
\usepackage{mathpazo}

\usepackage[LGR,T1]{fontenc}
\usepackage[latin9]{inputenc}
\usepackage{geometry}
\usepackage{color}
\usepackage{float}
\usepackage{amsmath}
\usepackage{amssymb}
\usepackage{cancel}
\usepackage{graphicx}

\makeatletter

\DeclareRobustCommand{\greektext}{%
  \fontencoding{LGR}\selectfont\def\encodingdefault{LGR}}
\DeclareRobustCommand{\textgreek}[1]{\leavevmode{\greektext #1}}

\providecommand{\tabularnewline}{\\}
\floatstyle{ruled}
\newfloat{algorithm}{tbp}{loa}
\providecommand{\algorithmname}{Algorithm}
\floatname{algorithm}{\protect\algorithmname}

\usepackage[version=3]{mhchem}
\usepackage{amsmath}
\usepackage{algorithm}
\usepackage{algpseudocode}
\usepackage{xcolor}
\usepackage{hyperref}

\makeatother

\usepackage{babel}
\begin{document}
\begin{frontmatter}
\title{A multiscale hybrid Maxwellian-Monte-Carlo Coulomb collision algorithm
for particle simulations}
\author[]{G. Chen\corref{cor1}}
\ead{gchen@lanl.gov}
\author[]{A. J. Stanier}
\author[]{L. Chac\'on}
\author[]{S. E. Anderson}
\author[]{B. Philip}
\cortext[cor1]{Corresponding author}
\address{Los Alamos National Laboratory, Los Alamos, NM 87545}
\begin{abstract}
Coulomb collisions in particle simulations for weakly coupled plasmas
are modeled by the Landau-Fokker-Planck equation, which is typically
solved by Monte-Carlo (MC) methods. One of the main disadvantages
of MC is the timestep accuracy constraint \textgreek{\textnu}\ensuremath{\Delta}t
\ensuremath{\ll} 1 to resolve the collision frequency \textgreek{\textnu}.
The constraint becomes extremely stringent for self-collisions in
the presence of high-charge state species and for inter-species collisions
with large mass disparities (such as present in Inertial Confinement
Fusion hohlraums), rendering long-time-scale simulations prohibitively
expensive or impractical. To overcome these difficulties, we explore
a hybrid Maxwellian-MC (HMMC) model for particle simulations. Specifically,
we devise a collisional algorithm that describes weakly collisional
species with particles, and highly collisional species and fluid components
with Maxwellians. We employ the Lemons method for particle-Maxwellian
collisions, enhanced with a more careful treatment of low-relative-speed
particles, and a five-moment model for Maxwellian-Maxwellian collisions.
Particle-particle binary collisions are dealt with classic Takizuka-Abe
MC, which we extend to accommodate arbitrary particle weights to deal
with large density disparities without compromising conservation properties.
HMMC is strictly conservative and significantly outperforms standard
MC methods in situations with large mass disparities among species
or large charge states, demonstrating orders of magnitude improvement
in computational efficiency. We will substantiate the accuracy and
performance of the proposed method with several examples of varying
complexity, including both zero-dimensional relaxation and one-dimensional
transport problems, the latter using a hybrid kinetic-ion/fluid-electron
model.
\end{abstract}
\end{frontmatter}

\section{Introduction}

In numerous laboratory and natural plasma applications, such as magnetic
and inertial fusion, space plasmas, and low-temperature plasma discharges,
there is a growing demand for reliable, long-time-scale kinetic simulations
that accurately account for Coulomb collisions. Our focus is on Coulomb
collisions in fully ionized, non-relativistic, weakly coupled plasmas,
which are typically described by the Landau-Fokker-Planck (LFP) equation
\citep{helander2005collisional}.

Conventional collisional approaches typically fall into two main categories:
particle-based and grid-based methods. Particle-based collision methods
are particularly attractive for implementation in particle-in-cell
(PIC) algorithms owing to their flexibility and widespread use. Among
these, particle-pairing methods utilizing Monte Carlo (MC) techniques
\citep{takizuka1977binary,nanbu1997theory} are most widely adopted.
The MC method offers several advantages, including a linear cost scaling
with the number of particles $N_{p}$, straightforward implementations,
automatic preservation of positivity, and the strict conservation
of mass, momentum, and energy. Integration with PIC methods are typically
achieved through first-order operator splitting, resulting in robust
PIC-MC simulation algorithms. \textcolor{red}{Higher-order splitting
methods, such as second-order Strang splitting, have been applied
in particle-based BGK-type collisions \citep{medaglia2023stochastic},
grid-based Landau-Fokker-Planck collisions \citep{crouseilles2004numerical},
and the DSMC method for neutral particle collisions \citep{hokazono2003time}.
However, to our knowledge, high-order operator splitting has yet to
be applied to the PIC-MC method for Coulomb collisions.}

Standard MC methods exhibit several limitations. As already mentioned,
their temporal convergence rate is typically slow ($\sim\sqrt{\Delta t}$)
\citep{wang2008particle,cohen2010time}. Their random character introduces
additional noise in already noisy particle simulations, exacerbating
errors. Another drawback of MC is the timestep constraint ($\nu\Delta t\ll1$)
to resolve all collision frequencies $\nu$ for accuracy \citep{wang2008particle,cohen2010time}.
The timestep constraint becomes particularly challenging for self-collisions
involving high-Z species (as $\nu\propto Z^{4}$), and for inter-species
collisions with significant mass disparities (as $\nu\propto m_{\alpha\beta}^{-1}$,
where the reduced mass $m_{\alpha\beta}$ will be approximately equal
to the smaller mass), rendering such simulations prohibitively expensive.
These challenges have motivated researchers to investigate alternative
approaches for disparate-mass collisions based on Langevin stochastic
models for particle-Maxwellian collisions \citep{sherlock2008monte,lemons2009small,higginson2022cartesian}.

\textcolor{red}{Asymptotic-preserving MC collisional schemes \citep{pareschi2000asymptotic,ren2014asymptotic,zhang2016asymptotic,dimarco2018asymptotic,crestetto2020new,mortier2020kinetic,lovbak2021multilevel,mortier2022multilevel}
have the potential of significant algorithmic acceleration, since
they allow large timesteps while still capturing the collisional equilibrium.
However, those studies consider non-Coulombian collision operators
such as Boltzmann/Bhatnagar-Gross-Krook (BGK), which do not trivially
extend to the Coulomb-collision case, governed by a Fokker-Planck-type
equation. }\textcolor{black}{Other authors have attempted to accelerate
MC for Coulomb collisions either by considering splitting the distribution
function into one or several Maxwellian components plus a kinetic
one \citep{larson2003coulomb,caflisch2008hybrid,lemons2009small,ricketson2014entropy,caflisch2016accelerated},
or by using multilevel MC methods \citep{rosin2014multilevel}. The
method most closely related to this study is the so-called hybrid
MC method, introduced by Caflisch \citep{caflisch2008hybrid} and
later improved by Ricketson \citep{ricketson2014entropy}. In these
approaches, the particle distribution function is decomposed into
a Maxwellian component (described analytically) plus a perturbation
(described with particles). As in our proposed algorithm, the computational
speedup follows from the fact that the Maxwellian self-collisions
are null, and no longer introduce a fast timescale. However, the actual
implementation of these approaches is cumbersome, as they explicitly
need to thermalize/dethermalize particles to keep the Maxwellian component
a Maxwellian during the collisional evolution.}

Other authors have proposed deterministic \textcolor{red}{structure-preserving
}particle collisional algorithms that sidestep MC altogether, avoiding
random noise \citep{carrillo2020particle,carrillo2021random,hirvijoki2021structure,carrillo2023convergence,bailo2024collisional}.
These methods employ a gradient-flow formulation to derive equations
of motion for the particles, which encode the collisional process
with high fidelity and preserve all collisional invariants (mass,
momentum, and energy) with implicit timestepping \citep{hirvijoki2021structure}
while guaranteeing an H-theorem for entropy generation. As a result,
they result in noise-free particle collision simulations. However,
these methods remain expensive, of $\mathcal{O}(N_{p}^{2})$, are
typically implemented explicitly in time (breaking energy conservation),
and so far require a velocity-space mesh for initialization due to
the particle regularization employed, which is subject to the curse
of dimensionality. Random batching \citep{carrillo2021random,bailo2024collisional}
can significantly ameliorate the computational scaling to $\mathcal{O}(N_{p}^{2}/R)$,
with $R$ the number of random batches, without formal loss of conservation
properties. However, accuracy considerations require that $R$ scale
sublinearly with $N_{p}$ \citep{carrillo2021random}, as well as
a sufficiently small timestep to recover collision statistics. Therefore,
despite significant acceleration, at present the approach scales superlinearly
with $N_{p}$ and is not multiscale in time, needing to resolve all
collisional time scales present.

Coulomb collisional methods utilizing a velocity grid have also been
extensively explored (see Ref. \citep{taitano2015mass} and references
therein for a good survey). Implicit time integration enables stepping
over stiff collisional timescales, with the potential of significant
efficiency gains while preserving all relevant conservation properties,
either directly with the LFP formulation \citep{yoon2014fokker,adams2017landau,adams2022landau,hager2016fully,hirvijoki2017conservative}
or the equivalent Rosenbluth-Fokker-Planck one \citep{taitano2015mass,taitano2016adaptive,taitano2018adaptive,Taitano2021,Taitano2021b}.
Conservation in the Landau form arises from symmetry, and is straightforward
to enforce, but the method is $\mathcal{O}(N^{2}),$ with $N$ the
total number of mesh points. In contrast, the Rosenbluth form can
be made $\mathcal{O}(N)$, but ensuring the conservation laws and
constraints for long-term accuracy necessitates specialized techniques,
which can be intricate \citep{taitano2015mass,taitano2016adaptive,taitano2018adaptive,Taitano2021,Taitano2021b}.
Acceleration techniques are indispensable to improve the convergence
rate of nonlinear iterative solvers required by the resulting nonlinear
algebraic systems, especially when dealing with stiff timescales,
a task that presents its own set of challenges \citep{Taitano2021b}.
In addition to the temporal integration challenges, attempting to
solve the LFP equation on a three-dimensional velocity grid quickly
leads to the so-called curse of dimensionality of tensor-product meshes
in high-dimensions when coupled with Vlasov's equation, rendering
it expensive for long-time applications on even today's fastest supercomputers.
Grid-based collisional methods can be combined with PIC simulations
\citep{hager2016fully} to avoid random MC noise. However, the frequent
interpolations between the phase-space grid and the particles can
lead to numerical diffusion \citep{denavit1972numerical} unless specialized
interpolation techniques are employed \citep{mollen2021implementation}.

This study proposes a hybrid Maxwellian-MC (HMMC) Coulomb-collision
approach in which colliding species can be described by either particles
or Maxwellians, the latter being an appropriate description under
the assumption of sufficiently fast self-collisions, or by ansatz
in hybrid fluid-kinetic models. As in earlier hybrid MC approaches
\citep{caflisch2008hybrid,ricketson2014entropy}, treating fast-colliding
species as a Maxwellian eliminates the fastest self-collisional timescales
of the system, resulting in significant algorithmic speedups. However,
unlike those earlier studies, our approach does not require particle
(de)thermalization because we only use the Maxwellian ansatz when
the collisionality regime warrants it. The approach considers all
types of collisions: among particles, among Maxwellians, and between
particles and Maxwellians. It scales as $\mathcal{O}(N_{p})$ and
strictly conserves all collisional invariants. Our approach offers
a completely general multiscale solution for arbitrary particle systems
undergoing Coulombian interactions, but is particularly suitable for
hybrid fluid-PIC algorithms (where electrons are commonly modeled
by a fluid species) and for systems featuring high-Z ion species because
the self-collision frequency scales as $Z^{4}$, and classical MC
methods quickly become expensive.

Since our approach considers coexisting particle and fluid species,
collisions between particles, between particles and Maxwellians, and
between Maxwellians need an appropriate treatment. Here, we employ
the Takizuka-Abe (TA) MC method \citep{takizuka1977binary} for particle-particle
binary collisions\textcolor{red}{{} (alternatively, Nanbu-Bobylev's
methods \citep{nanbu1997theory,bobylev2000theory} can be used),}
the Lemons method \citep{lemons2009small} for particle-Maxwellian
collisions, and the 5-moment model for collisions between Maxwellian
species \citep{burgers1969flow,echim2011review}. Importantly, we
enhance the Lemons method with a more careful treatment of low-relative-speed
particles, without which the method may either produce erroneous results
or be very inefficient \citep{cohen2010time}. Moreover, we also extend
the standard TA method to accommodate arbitrary particle weights without
compromising conservation properties. This can be particularly useful
for collisions between species with large density disparities. Previous
studies on variable-weight Coulomb collision schemes were often developed
in an ad-hoc manner, with a primary focus on conserving momentum and
energy on average \citep{miller1994coulomb,nanbu1998weighted,sentoku2008numerical,higginson2020corrected}.
However, there are notable exceptions to this trend. Shanny et al.
derived their scheme specifically for simulating electrons in electron-ion
collisions\textcolor{red}{{} within the Lorentz model} \citep{shanny1967one}.
Tanaka et al., on the other hand, introduced a correction step aimed
at achieving exact momentum and energy conservation. However, they
did not provide a detailed derivation of their proposed particle pairing
scheme \citep{tanaka2018coulomb}. The proposed method is strictly
conservative in mass, momentum, and energy, and as we will show significantly
outperforms classical MC methods, with orders of magnitude improvement
vs. TA in computational efficiency. We will substantiate the accuracy
and performance of the proposed method with several examples of varying
complexity, including both relaxation and transport problems.

This paper is organized as follows. Section \ref{sec:Methods} introduces
the three models utilized in this study for particle-particle, particle-Maxwellian,
and Maxwellian-Maxwellian collisions. In Sec. \ref{sec:Numerical-experiments},
HMMC is demonstrated with various difficult benchmarks, including
zero-dimensional multi-species relaxation and one-dimensional multi-species
transport in an Inertial Confinement Fusion (ICF) hohlraum-like environment.
Finally, we summarize the study and conclude in Sec. \ref{sec:conclusions}.

\section{Methodology}

\label{sec:Methods}

We aim for a versatile multiscale collisional particle algorithm that
allows any highly collisional species to be represented as a Maxwellian
during the collision process. This, in turn, requires suitable algorithmic
solutions to deal with particle-particle, particle-Maxwellian, and
Maxwellian-Maxwellian collisions.

For particle-particle collisions, we consider the well-known TA algorithm.
However, the conventional TA model encounters challenges when dealing
with collisions between species with significant density disparities.
This often occurs in scenarios where a species is moving into an empty
space or when the system contains a minority species. The TA algorithm
assumes a uniform particle weight between collision pairs, which can
lead to either an excessive number of particles for the high-density
species or an inadequate number for the low-density ones, resulting
in efficiency issues or enhanced noise, respectively. To circumvent
these difficulties, non-uniform weight schemes are preferred \citep{miller1994coulomb,nanbu1998weighted,sentoku2008numerical,higginson2020corrected}.
In this study, we devised a new particle pairing scheme for non-uniform
TA collisions, incorporating a correction step to ensure exact conservation
of momentum and energy. This approach addresses the shortcomings of
the conventional TA model and enhances the accuracy and efficiency
of particle-particle collision treatments in strongly non-uniform
plasma systems.

Particle-Maxwellian collisions are dealt with an improved version
of the Lemons algorithm \citep{lemons2009small}. The Lemons method,
formulated in spherical coordinates, has proven to be more efficient
than Cartesian Langevin equations \citep{cohen2010time} in scenarios
where a light particle species collides with a heavy Maxwellian species,
making it a more palatable choice. Additionally, we have fixed a failure
mode that we uncovered in the treatment of low-relative-speed particles
in the standard Lemons algorithm that renders the method accurate
for virtually any particle-Maxwellian interaction.

Maxwellian-Maxwellian collisions are modeled with a five-moment model
describing the evolution of the Maxwellian moments\textcolor{red}{{}
(i.e., density, bulk velocity, and temperature)}. The five-moment
equations are derived exactly by moment integration of the LFP equation
when assuming the Maxwellian remains so dynamically, and can be shown
to preserve all conservation properties exactly \citep{burgers1969flow,echim2011review}.
The five-moment model eliminates self-collisional timescales from
the formulation, which is particularly advantageous when dealing with
stiff self-colliding species.

In the following sub-sections, we discuss first the Maxwellian-Maxwellian
collision method, next the improved Lemons method, and last the extended
TA method. We also discuss the special case of hybrid kinetic-ion/fluid-electron
algorithms \citep{winske2023hybrid}, which is a common model of choice
for various applications, and demands a specialized treatment of the
fluid electron system to ensure strict conservation properties of
collisional invariants \citep{sherlock2008monte}. Finally, we briefly
comment on the algorithmic orchestration such that conservation properties
are preserved during the whole collisional step.

\subsection{Maxwellian-Maxwellian collisions: 5-moment model}

A three-dimensional Maxwellian distribution function is fully determined
by its first five moments ($n,\boldsymbol{u},T$), and may be written
as:
\[
\varphi(\mathbf{v})=n\left(\frac{m}{2\pi kT}\right)^{3/2}\exp\left(-\frac{m(\mathbf{v}-\mathbf{u})^{2}}{2kT}\right),
\]
where the five moments are defined as:
\begin{align*}
n & =\int\varphi(\mathbf{v})d\mathbf{v},\\
\mathbf{u} & =\frac{1}{n}\int\varphi(\mathbf{v})\mathbf{v}d\mathbf{v},\\
T & =\frac{m}{3kn}\int\varphi(\mathbf{v})(\mathbf{v}-\mathbf{u})^{2}d\mathbf{v},
\end{align*}
\textcolor{red}{which are number density, three components of bulk
velocity, and temperature, respectively.} Here, $m$ is mass and $k$
is the Boltzmann constant. If we further assume that the Maxwellian
is preserved during the collision process, then the evolution equations
for the moments undergoing Coulomb collisions can be exactly formulated.
This is done by taking the moments of the Landau-Fokker-Planck equation,
resulting in the following equations of motion for the moments of
the Maxwellian distribution \citep{burgers1969flow,echim2011review}:
\begin{align}
\frac{dn_{\alpha}}{dt} & =0,\label{eq:dndt}\\
\frac{d\mathbf{u}_{\alpha}}{dt} & =\sum_{\beta}\nu_{\alpha\beta}(\mathbf{u}_{\beta}-\mathbf{u}_{\alpha})\Phi_{\alpha\beta},\label{eq:dudt}\\
\frac{3}{2}k\frac{dT_{\alpha}}{dt} & =\sum_{\beta}\frac{m_{\alpha}}{m_{\alpha}+m_{\beta}}\nu_{\alpha\beta}\left[3k(T_{\beta}-T_{\alpha})\Psi_{\alpha\beta}+m_{\beta}(\mathbf{u}_{\beta}-\mathbf{u}_{\alpha})^{2}\Phi_{\alpha\beta}\right],\label{eq:dTdt}
\end{align}
where 
\begin{align}
\nu_{\alpha\beta} & =\frac{1}{3}\frac{n_{\beta}m_{\beta}}{m_{\alpha}+m_{\beta}}\left(\frac{2\pi kT_{\alpha\beta}}{m_{\alpha\beta}}\right)^{-3/2}\frac{e_{\alpha}^{2}e_{\beta}^{2}}{\varepsilon_{0}^{2}m_{\alpha\beta}^{2}}\mathrm{ln}\Lambda,\label{eq:nu_ab_M}\\
m_{\alpha\beta} & =\frac{m_{\alpha}m_{\beta}}{m_{\alpha}+m_{\beta}},\nonumber \\
T_{\alpha\beta} & =\frac{m_{\beta}T_{\alpha}+m_{\alpha}T_{\beta}}{m_{\alpha}+m_{\beta}},\nonumber \\
\Phi_{\alpha\beta} & =\frac{3}{2u_{\alpha\beta}^{2}}\left(\frac{\sqrt{\pi}}{2}\frac{\phi(u_{\alpha\beta})}{u_{\alpha\beta}}-\exp(-u_{\alpha\beta}^{2})\right),\nonumber \\
\Psi_{\alpha\beta} & =\exp(-u_{\alpha\beta}^{2}),\nonumber \\
u_{\alpha\beta} & =\frac{|\mathbf{u}_{\alpha}-\mathbf{u}_{\beta}|}{\sqrt{2kT_{\alpha\beta}/m_{\alpha\beta}}},\nonumber 
\end{align}
and $\phi(x)=\frac{2}{\sqrt{\pi}}\int_{0}^{x}e^{-y^{2}}dy$ is the
error function. To facilitate a conservative discretization of the
energy equation, we reformulate Eq. \ref{eq:dTdt} in terms of the
energy $\varepsilon_{\alpha}=\frac{1}{2}m_{\alpha}u_{\alpha}^{2}+\frac{3}{2}kT_{\alpha}$
as: 
\begin{equation}
\frac{d\varepsilon_{\alpha}}{dt}=\sum_{\beta}\frac{m_{\alpha}\mathbf{u}_{\alpha}+m_{\beta}\mathbf{u}_{\beta}}{m_{\alpha}+m_{\beta}}m_{\alpha}\nu_{\alpha\beta}(\mathbf{u}_{\beta}-\mathbf{u}_{\alpha})\Phi_{\alpha\beta}+\frac{m_{\alpha}3k(T_{\beta}-T_{\alpha})}{m_{\alpha}+m_{\beta}}\nu_{\alpha\beta}\Psi_{\alpha\beta}.\label{eq:dEdt}
\end{equation}

\textcolor{red}{Since the five-moment description is intended to describe
fast collisional processes,} we employ a \textcolor{red}{backward
Euler} discretization of Eqs. \ref{eq:dudt}-\ref{eq:dEdt} \textcolor{red}{to
ensure integration stability and asymptotic accuracy for large timesteps,
namely}:
\begin{align}
\frac{\mathbf{u}_{\alpha}^{n+1}-\mathbf{u}_{\alpha}^{n}}{\Delta t} & =\sum_{\beta}\left(\nu_{\alpha\beta}(\mathbf{u}_{\beta}-\mathbf{u}_{\alpha})\Phi_{\alpha\beta}\right)^{{\color{red}n+1}},\label{eq:dudt-Maxwell}\\
\frac{\varepsilon_{\alpha}^{n+1}-\varepsilon_{\alpha}^{n}}{\Delta t} & =\sum_{\beta}\left(\frac{m_{\alpha}\mathbf{u}_{\alpha}+m_{\beta}\mathbf{u}_{\beta}}{m_{\alpha}+m_{\beta}}m_{\alpha}\nu_{\alpha\beta}(\mathbf{u}_{\beta}-\mathbf{u}_{\alpha})\Phi_{\alpha\beta}+\frac{m_{\alpha}3k(T_{\beta}-T_{\alpha})}{m_{\alpha}+m_{\beta}}\nu_{\alpha\beta}\Psi_{\alpha\beta}\right)^{{\color{red}n+1}},\label{eq:dedt-Maxwell}
\end{align}
with the right-hand-side computed \textcolor{red}{at $n+1$}. It is
straightforward to see that conservation of momentum and energy follow
by symmetry:
\begin{align*}
\sum_{\alpha}(\mathbf{u}_{\alpha}^{n+1}-\mathbf{u}_{\alpha}^{n}) & =0,\\
\sum_{\alpha}(\varepsilon_{\alpha}^{n+1}-\varepsilon_{\alpha}^{n}) & =0.
\end{align*}
Since the formulae were first derived by Burgers \citep{burgers1969flow},
hereafter we call it the Burgers method.\textcolor{red}{{} The solution
process of Eqs. \ref{eq:dudt-Maxwell}, \ref{eq:dedt-Maxwell} is
as follows. We consider two-species interactions at a time. The resulting
nonlinear system, $\mathbf{F}(\mathbf{x})=0$ with $\mathbf{x}=[u_{\alpha1},u_{\alpha2},u_{\alpha3},\varepsilon_{\alpha},u_{\beta1},u_{\beta2},u_{\beta3},\varepsilon_{\beta}]^{T}$,
is solved iteratively using a quasi-Newton method. In each iteration,
an approximate Jacobian system $A\Delta\mathbf{x}_{k}=-\mathbf{F}(\mathbf{x}_{k})$
is solved, with $A$ a suitable approximation to the Jacobian matrix,
and the solution is updated as $\mathbf{x}_{k+1}=\mathbf{x}_{k}+\Delta\mathbf{x}_{k}$.
Here, $A$ is obtained by Picard-linearizing the collision frequencies
and Chandrasekhar functions to the previous iteration. The Jacobian
equation for the update is solved by the LU decomposition method.
Iterations are terminated when the relative error is smaller than
$10^{-8}.$}

\subsection{Particle-Maxwellian collisions: improved Lemons method}

\label{subsec:lemons}

The Lemons algorithm \citep{lemons2009small} is a \textquotedbl particle-moment\textquotedbl{}
collision algorithm designed for Coulomb collisions. This method involves
a set of stochastic differential equations (SDE) that account for
particle collisions with a \textquotedbl fluid\textquotedbl{} species
characterized by a Maxwellian distribution. During each timestep,
particles are scattered once with a \textquotedbl fluid\textquotedbl{}
species within a cell. The scattering process is integrated using
finite-difference solutions to stochastic differential equations that
incorporate Spitzer's velocity-space diffusion coefficients. A scattering
event is characterized by changes in the deflective angle $d\theta$,
azimuthal angle $d\phi$ , and relative speed $d\omega$ (note that
$\omega=|\mathbf{v}_{t}-\mathbf{v}_{f}|$, with $\mathbf{v}_{t}$
the test particle velocity and $\mathbf{v}_{f}$ the fluid drift velocity;
the angles ($\theta,\phi$) are defined with $\omega$ aligned with
the $z$-coordinate). The time evolution of those quantities is governed
by: 
\begin{align}
d\theta & =\sqrt{2\gamma dt}N_{\theta}(0,1),\label{eq:SDE-dtheta}\\
d\phi & =2\pi U_{\phi}(0,1),\label{eq:SDE-dphi}\\
d\omega & =-\beta\omega dt+\sqrt{\delta^{2}dt}N_{\omega}(0,1),\label{eq:SDE-domega}
\end{align}
where $N_{\theta}(0,1)$ and $N_{\omega}(0,1)$ are normal random
variables with zero mean and unit variance, $U_{\phi}(0,1)$ is a
uniform random variable distributed between zero and one. The coefficients
$\gamma$, $\beta$ and $\delta^{2}$ are found from matching Chandrasekhar\textquoteright s
classical formulas \citep{spitzer2006physics}: 
\begin{align}
\frac{d}{dt}\left\langle v_{z}\right\rangle  & =-l_{f}^{2}A_{D}(1+\frac{m_{t}}{m_{f}})G(\hat{\omega}),\label{eq:dynamic-friction}\\
\frac{d}{dt}\left(\left\langle v_{z}^{2}\right\rangle -\left\langle v_{z}\right\rangle ^{2}\right) & =\frac{A_{D}}{\omega}G(\hat{\omega}),\label{eq:diffusion-coef-z}\\
\frac{d}{dt}\left(\left\langle v_{\perp}^{2}\right\rangle -\left\langle v_{\perp}\right\rangle ^{2}\right) & =\frac{A_{D}}{\omega}[\phi(\hat{\omega})-G(\hat{\omega})],\label{eq:diffusion-coef-perp}
\end{align}
where $\left\langle v_{z}\right\rangle $ denotes expectation value
over the particle distribution, the subscript $z$ denotes the direction
of the relative velocity, the subscript $\perp$ denotes direction
perpendicular to the relative velocity, subscripts $f$ and $t$ denote
field and test particles respectively, $A_{D}=8\pi n_{f}q_{t}^{2}q_{f}^{2}\ln\Lambda/m_{t}^{2}$,
and $\ln\Lambda$ the Coulomb logarithm, $\hat{\omega}=\omega l_{f}$
with $l_{f}=\sqrt{m_{f}/2T_{f}}=1/v_{th,f}$ the inverse thermal velocity
of the fluid, $\phi(x)$ is the error function, and: 
\begin{align}
G(x) & =\frac{\phi(x)-x\phi^{\prime}(x)}{2x^{2}},\label{eq:Chandrasekhar-function}
\end{align}
with $\phi^{\prime}(x)=d\phi(x)/dx$. By expressing the dynamic friction
and diffusion coefficients through Eqs. \ref{eq:SDE-dtheta}-\ref{eq:SDE-domega}
and comparing those with Eqs. \ref{eq:dynamic-friction}-\ref{eq:diffusion-coef-perp},
one finds that: 
\begin{align}
\gamma & =\frac{A_{D}}{2\omega^{3}}[\phi(\hat{\omega})-G(\hat{\omega})],\label{eq:expression-gamma}\\
\beta & =\frac{A_{D}}{2\omega^{3}}\left\{ G(\hat{\omega})\left[\left(1+\frac{m_{t}}{m_{f}}\right)2\hat{\omega}^{2}+1\right]-\phi(\hat{\omega})\right\} ,\label{eq:expression-beta}\\
\delta^{2} & =\frac{A_{D}G(\hat{\omega})}{\omega}.\label{eq:expression-delta2}
\end{align}

We discretize Eqs. \ref{eq:SDE-dtheta}-\ref{eq:SDE-domega} in time
\textcolor{red}{using a predictor-corrector method} as:
\begin{align}
\theta^{n+1} & =\theta^{n}+\sqrt{2\gamma^{n\mathbin{\color{red}+}{\color{red}1/2}}\Delta t}N_{\theta}(0,1),\label{eq:SDE-dtheta-1}\\
\phi^{n+1} & =\phi^{n}+2\pi U_{\phi}(0,1),\label{eq:SDE-dphi-1}\\
\omega^{n+1} & =e^{-\beta^{n\mathbin{\color{red}+}{\color{red}1/2}}\Delta t}\omega^{n}+\sqrt{(\delta^{n\mathbin{\color{red}+}{\color{red}1/2}})^{2}\Delta t}N_{\omega}(0,1)+\frac{1}{2}\delta^{n\mathbin{\color{red}+}{\color{red}1/2}}\delta^{\prime n\mathbin{\color{red}+}{\color{red}1/2}}\Delta t\left(N_{\omega}(0,1)^{2}-1\right),\label{eq:SDE-domega-1}
\end{align}
where $\gamma^{n\mathbin{\color{red}+}{\color{red}1/2}}$, $\beta^{n\mathbin{\color{red}+}{\color{red}1/2}}$,
and $\delta^{n\mathbin{\color{red}+}{\color{red}1/2}}$ are obtained
by evaluating Eqs. \ref{eq:expression-gamma}-\ref{eq:expression-delta2}
using the timestep $n\mathbin{\color{red}+}{\color{red}1/2}$ quantities
\textcolor{red}{obtained as indicated below}, and we have employed
the Milstein scheme \citep{mil1975approximate} for the speed ($\omega$)
update, which is a first-order strong-convergence temporal scheme.
In Eq. \ref{eq:SDE-domega-1},
\begin{equation}
\delta\delta^{\prime}=-\frac{A_{D}}{4\omega^{3/2}}\left(\phi''(\hat{\omega})+6G(\hat{\omega})\right)\label{eq:2nd-order-term-Milstein}
\end{equation}
with $\phi''(x)=-\frac{4x}{\sqrt{\pi}}e^{-x^{2}}$, and we have analytically
integrated the deterministic (friction) term. \textcolor{red}{Note
that} the same normal random number\textcolor{red}{{} is used for $N_{\omega}(0,1)$}
in Eq. \ref{eq:SDE-domega-1}\textcolor{red}{{} to ensure that the mean
square deviation, $\left\langle \omega^{2}\right\rangle $, retains
second-order accuracy in the Ito-Taylor expansion }\citep{mil1975approximate}.\textcolor{red}{{}
The $n\mathbin{\color{red}+}{\color{red}1/2}$ quantities in Eqs.
\ref{eq:SDE-dtheta-1} to \ref{eq:SDE-domega-1} are predicted using
an explicit half-timestep ($\Delta t/2$) as follows:
\begin{align}
\theta^{n+1/2} & =\theta^{n}+\sqrt{2\gamma^{n}\Delta t/2}N_{\theta}(0,1),\label{eq:SDE-dtheta-1-1}\\
\phi^{n+1/2} & =\phi^{n}+2\pi U_{\phi}(0,1),\label{eq:SDE-dphi-1-1}\\
\omega^{n+1/2} & =e^{-\beta^{n}\Delta t/2}\omega^{n}+\sqrt{(\delta^{n})^{2}\Delta t/2}N_{\omega}(0,1)+\frac{1}{2}\delta^{n}\delta^{\prime n}\Delta t/2\left(N_{\omega}(0,1)^{2}-1\right).\label{eq:SDE-domega-1-1}
\end{align}
}

\textcolor{red}{Once Eqs. \ref{eq:SDE-dtheta-1}-\ref{eq:SDE-domega-1}
are solved, the collisional evolution of the background Maxwellian
distribution of species $\alpha$, characterized by ($n_{\alpha},\boldsymbol{u}_{\alpha},T_{\alpha}$),
resulting from Lemons collisions with particles of species $\beta$
is determined through conservation of momentum and energy as \citep{sherlock2008monte}:
\begin{align}
m_{\alpha}n_{\alpha}(\mathbf{u}_{\alpha}^{n+1/2}-\mathbf{u}_{\alpha}^{n}) & =\mathbf{M}_{\beta}^{n+1/2}-\mathbf{M}_{\beta}^{n},\label{eq:mom-conserv-Lemons}\\
m_{\alpha}n_{\alpha}(\varepsilon_{\alpha}^{n+1/2}-\varepsilon_{\alpha}^{n}) & =E_{\beta}^{n+1/2}-E_{\beta}^{n},\label{eq:energy-conserv-Lemons}
\end{align}
where $\mathbf{M}_{\beta}=\sum_{\beta}m_{\beta}\mathbf{v}_{\beta}$
and $E_{\beta}=\frac{1}{2}\sum_{\beta}m_{\beta}v_{\beta}^{2}$ are
the total momentum and energy of the species $\beta$, and as before:
\begin{equation}
\varepsilon_{\alpha}=\frac{1}{2}m_{\alpha}u_{\alpha}^{2}+\frac{3}{2}kT_{\alpha}.\label{eq:En-to-T}
\end{equation}
}

The exponential integration of the deterministic (friction) term in
Eq. \ref{eq:SDE-domega-1} prevents negative values from the deterministic
term for any timestep, and allows relatively large timesteps without
resulting in negative values from the random term. If that happens
and $\omega^{n+1}<0$ for a given particle, we invoke the low-relative-speed
particle treatment, which we describe next.

\subsubsection{Low-relative-speed particle treatment in Lemons}

We note that $\gamma$, $\beta$ and $\delta^{2}$ are singular as
$\omega\rightarrow0$. The mollification strategy for the singularity
provided by Lemons in Ref. \citep{lemons2009small} is to Taylor-expand
for sufficiently small $\omega$ to find the leading-order terms in
$\beta$ (from Eq. \ref{eq:expression-beta}), which are then plugged
into Eq. \ref{eq:SDE-domega}. Neglecting the second diffusion term
in Eq. \ref{eq:SDE-domega}, we find: 
\begin{equation}
d\omega=\frac{2A_{D}l_{f}dt}{3\sqrt{\pi}\omega},\label{eq:domega-1st-order}
\end{equation}
 which can be used to update $\omega$ as:
\begin{equation}
\omega(t+\Delta t)=\sqrt{\omega^{2}+\frac{4}{3\sqrt{\pi}}A_{D}l_{f}\Delta t}.\label{eq:omega-update-1st-order}
\end{equation}
As noted in Ref. \citep{lemons2009small}, Eq. \ref{eq:omega-update-1st-order}
differs from the deterministic part in Eq. \ref{eq:SDE-domega} when:
\begin{equation}
\omega^{2}\leq\frac{4}{3\sqrt{\pi}}A_{D}l_{f}\Delta t,\label{eq:small-omega-condition}
\end{equation}
which defines when $\omega$ is sufficiently small.

However, the small-$\omega$ limit holds true only under certain conditions:
1) the argument of the error and Chandrasekhar functions (i.e., $\hat{\omega}\equiv\omega l_{f}=\omega/v_{th,f}$)
must also be small; and 2) it must also be small enough so that the
physical response of the particle is acceleration (not deceleration).
Condition 1) may become invalid when $l_{f}$ is large (i.e., for
small fluid thermal velocity). But more importantly, Equation \ref{eq:omega-update-1st-order}
always accelerates the particle, which is unphysical and eventually
breaks the smallness of $\omega$ regardless of its initial value.

To fix the first inconsistency, we draw inspiration from the classical
TA MC algorithm \citep{takizuka1977binary}, which proposes that scattering
becomes isotropic when $\sigma^{2}=\frac{e_{\alpha}^{2}e_{\beta}^{2}n_{L}\ln\Lambda}{8\pi\epsilon_{0}m_{\alpha\beta}^{2}\omega^{3}}\Delta t>1$
(see e.g. Eq. \ref{eq:TA-f(tantheta)} in the next section). This
insight agrees with MC simulations conducted in Ref. \citep{nanbu1997theory},
which demonstrate that isotropic scattering occurs when $s\equiv2\sigma^{2}\gtrsim4$,
which is the criterion we adopt in this study. This condition defines
a smallness threshold for $\omega$, given by:
\[
\omega^{3}\lesssim\frac{e_{\alpha}^{2}e_{\beta}^{2}n_{L}\ln\Lambda}{16\pi\epsilon_{0}m_{\alpha\beta}^{2}}\Delta t,
\]
below which we perform isotropic scattering, thereby avoiding the
singularity of Eq. \ref{eq:SDE-domega}.

To prevent the particles from always accelerating in the small-$\omega$
regime (as would be dictated by Eq. \ref{eq:omega-update-1st-order}),
we derive a more appropriate formula for determining the relative
velocity by considering the averaged $\omega^{2}$. We may write:
\begin{align*}
d\omega^{2} & =(\omega+d\omega)^{2}-\omega^{2}\\
 & =2\omega d\omega+(d\omega)^{2}\\
 & =2\omega\left(-\beta\omega dt+\sqrt{\delta^{2}dt}N_{\omega}(0,1)\right)+\left(-\beta\omega dt+\sqrt{\delta^{2}dt}N_{\omega}(0,1)\right)^{2}\\
 & =-2\beta\omega^{2}dt+2\omega\sqrt{\delta^{2}dt}N_{\omega}(0,1)+\delta^{2}dt[N_{\omega}(0,1)]^{2},
\end{align*}
where we have neglected higher-order $dt$-terms. By taking the average
of the above equation, we get 
\begin{align}
<d\omega^{2}> & =-2\beta\omega^{2}dt+\delta^{2}dt\nonumber \\
 & =-2A_{D}l_{f}\left(\frac{m_{t}}{m_{f}}\hat{\omega}G(\hat{\omega})-\frac{e^{-\hat{\omega}^{2}}}{\sqrt{\pi}}\right)dt.
\end{align}
Since low-relative-speed particles will experience a large number
of small-angle collisions, the accumulated effect can be modeled as
isotropic scattering \citep{nanbu1997theory}. By approximating $<d\omega^{2}>/dt\approx d\omega^{2}/dt$,
we aim at counting the average energy exchange between the particle
and the field distribution, and finally get:

\begin{equation}
\frac{d\omega^{2}}{dt}=-2A_{D}l_{f}\left(\frac{m_{t}}{m_{f}}\hat{\omega}G(\hat{\omega})-\frac{e^{-\hat{\omega}^{2}}}{\sqrt{\pi}}\right)=S.\label{eq:energy-equation}
\end{equation}
In this study, we solve Eq. \ref{eq:energy-equation} with a predictor-corrector
scheme, written as:
\begin{align}
\frac{(\omega^{2})^{*}-(\omega^{2})^{n}}{\Delta t} & =S^{n},\label{eq:energy-equation-predictor}\\
\frac{(\omega^{2})^{**}-(\omega^{2})^{*}}{\Delta t} & =S^{*},\label{eq:energy-equation-corrector}
\end{align}
where the superscript {*} indicates the predicted solution using the
$S$ (the right hand side of Eq. \ref{eq:energy-equation}) evaluated
at time level $n$, while the corrected solution (indicated by the
superscript~{*}{*}) is obtained from $S$ evaluated using $\omega^{*}$.
The final solution is found as the temporal average of the two: 
\begin{equation}
\omega^{n+1}=\frac{\omega^{*}+\omega^{**}}{2}.
\end{equation}
After the collision process, the coordinate reference is changed back
from the $z$-aligned $\omega$ to the lab frame \citep{lemons2009small}.

\subsection{Particle-particle collisions with variable weights: extended TA method}

\label{subsec:extended-TA}

In this section, we present a derivation of the TA scheme. Our derivation
follows a similar approach to the one undertaken by Shanny et al.
\citep{shanny1967one} for the electron species in electron-ion collisions.
While the TA method originally extended Shanny's work to multi-component
plasmas, existing literature lacks a detailed algorithmic derivation.
However, we emphasize the significance of this derivation as it sheds
light on the rationale behind particle pairing choices (which will
inform our method for the variable weight scheme) and elucidates the
necessary conditions for ensuring the method's accuracy.

The TA method \citep{takizuka1977binary} is a Monte-Carlo method
that employs a particle-pairing scheme for simulating Coulomb collisions
of the Landau-Fokker-Planck operator. In its original form, a cumulative
scattering event (between particles of species \ensuremath{\alpha}
and \ensuremath{\beta} with a relative velocity \ensuremath{u}) has
a deflective angle \ensuremath{\theta}, the tangent of which satisfies
a normal distribution:
\begin{equation}
f(\delta)=\frac{1}{\sigma\sqrt{2\pi}}e^{-\frac{\delta^{2}}{2\sigma^{2}}},\label{eq:TA-f(tantheta)}
\end{equation}
where $\delta=\tan(\Theta/2)$ and the variance $\sigma^{2}$ is given
by:
\begin{equation}
\sigma^{2}=\frac{q_{\alpha}^{2}q_{\beta}^{2}n_{L}\ln\Lambda}{8\pi\epsilon_{0}m_{\alpha\beta}^{2}u^{3}}\Delta t.\label{eq:TA-var}
\end{equation}
Here $e_{\alpha}$, $e_{\beta}$ are the charge number of species
$\alpha$ and $\beta$ respectively, $n_{L}$ is the lower number
density between $n_{\alpha}$ and $n_{\beta}$, $\ln\Lambda$ is the
Coulomb logarithm, $\epsilon_{0}$ is the vacuum permittivity, $m_{\alpha\beta}=m_{\alpha}m_{\beta}/(m_{\alpha}+m_{\beta})$,
$u=|\mathbf{u}|$, and $\mathbf{u}=\mathbf{v}_{\alpha}-\mathbf{v}_{\beta}$
is the relative velocity of the paired particles. In a center-of-mass
coordinate system where the relative velocity is $(0,0,u)$ (see Eq.
2 of Ref. \citep{takizuka1977binary}), the change of velocity can
be expressed as:
\begin{align}
\Delta u_{x} & =u\sin(\Theta)\cos(\Phi),\nonumber \\
\Delta u_{y} & =u\sin(\Theta)\sin(\Phi),\label{eq:Delta-u-TA}\\
\Delta u_{z} & =-u(1-\cos\Theta),\nonumber 
\end{align}
where $\Theta$ is a normal random number sampled from the PDF of
Eq. \ref{eq:TA-f(tantheta)}, and $\Phi$ is a uniform random number
in $(0,2\pi)$. Based on the velocity change, one can derive the LFP
equation, which may be written as:
\[
\frac{df_{\alpha}}{dt}=-\nabla_{v}\cdot\mathbf{j}_{\alpha},
\]
where the components of the collisional flux of species $\alpha$
in velocity space are given by:
\[
j_{\alpha,i}=\frac{1}{m_{\alpha}}F_{i}^{\alpha/\beta}f_{\alpha}-D_{ik}^{\alpha/\beta}\frac{\partial f_{\alpha}}{\partial v_{k}},
\]
with:
\begin{align*}
\frac{1}{m_{\alpha}}F_{i}^{\alpha/\beta} & =\left\langle \Delta v_{i}\right\rangle ^{\alpha/\beta}-\frac{1}{2}\frac{\partial}{\partial v_{k}}\left\langle \Delta v_{i}\Delta v_{k}\right\rangle ^{\alpha/\beta},\\
D_{ik}^{\alpha/\beta} & =\frac{1}{2}\left\langle \Delta v_{i}\Delta v_{k}\right\rangle ^{\alpha/\beta}.
\end{align*}
Here, $\left\langle \Delta v\right\rangle $ denotes the ensemble
average of the rate of change of velocity \citep{trubnikov1965particle},
defined as:
\[
\left\langle \Delta v\right\rangle =\frac{1}{\Delta t}\int F(v,\Delta v)\Delta vd\Delta v,
\]
where $F(v,\Delta v)$ is the probability that the velocity of a particle
changes from $v$ to $v+\Delta v$ as a result of collisions in the
time $\Delta t$. Since the velocity change is given as $\Delta\mathbf{v}_{\alpha}=\frac{m_{\alpha\beta}}{m_{\alpha}}\Delta\mathbf{u}$,
consistency with the LFP equation demands: 
\begin{align}
\left\langle \Delta v_{x}\right\rangle ^{\alpha/\beta} & =\left\langle \Delta v_{y}\right\rangle ^{\alpha/\beta}=0=\left\langle \Delta v_{x}\right\rangle _{LFP}^{\alpha/\beta}=\left\langle \Delta v_{y}\right\rangle _{LFP}^{\alpha/\beta},\\
\left\langle \Delta v_{z}\right\rangle ^{\alpha/\beta} & =\frac{m_{\alpha\beta}}{m_{\alpha}}\left\langle \Delta u_{z}\right\rangle =-\frac{m_{\alpha\beta}}{m_{\alpha}}u\left\langle \frac{2\tan^{2}\Theta/2}{1+\tan^{2}\Theta/2}\right\rangle \approx-\frac{e_{\alpha}^{2}e_{\beta}^{2}n_{\beta}\ln\Lambda}{4\pi\epsilon_{0}m_{\alpha\beta}m_{\alpha}u^{2}}=\left\langle \Delta v_{z}\right\rangle _{LFP}^{\alpha/\beta},\label{eq:<dvz>}\\
\left\langle \Delta v_{x}\Delta v_{x}\right\rangle ^{\alpha/\beta} & =\left(\frac{m_{\alpha\beta}}{m_{\alpha}}\right)^{2}u^{2}\left\langle \frac{4\tan^{2}\Theta/2}{(1+\tan^{2}\Theta/2)^{2}}\right\rangle \left\langle \cos^{2}\theta\right\rangle \approx\frac{e_{\alpha}^{2}e_{\beta}^{2}n_{\beta}\ln\Lambda}{4\pi\epsilon_{0}m_{\alpha}^{2}u}=\left\langle \Delta v_{x}\Delta v_{x}\right\rangle _{LFP}^{\alpha/\beta},\label{eq:<dvx2>}\\
\left\langle \Delta v_{y}\Delta v_{y}\right\rangle ^{\alpha/\beta} & =\left(\frac{m_{\alpha\beta}}{m_{\alpha}}\right)^{2}u^{2}\left\langle \frac{4\tan^{2}\Theta/2}{(1+\tan^{2}\Theta/2)^{2}}\right\rangle \left\langle \sin^{2}\theta\right\rangle \approx\frac{e_{\alpha}^{2}e_{\beta}^{2}n_{\beta}\ln\Lambda}{4\pi\epsilon_{0}m_{\alpha}^{2}u}=\left\langle \Delta v_{y}\Delta v_{y}\right\rangle _{LFP}^{\alpha/\beta},\label{eq:<dvy2>}\\
\left\langle \Delta v_{z}\Delta v_{z}\right\rangle ^{\alpha/\beta} & =\left(\frac{m_{\alpha\beta}}{m_{\alpha}}\right)^{2}u^{2}\left\langle \frac{4\tan^{4}\Theta/2}{(1+\tan^{2}\Theta/2)^{2}}\right\rangle \approx0=\left\langle \Delta v_{y}\Delta v_{y}\right\rangle _{LFK}^{\alpha/\beta},\label{eq:<dvz2>}\\
\left\langle \Delta v_{x}\Delta v_{y}\right\rangle ^{\alpha/\beta} & =\left\langle \Delta v_{x}\Delta v_{z}\right\rangle ^{\alpha/\beta}=\left\langle \Delta v_{y}\Delta v_{z}\right\rangle ^{\alpha/\beta}=0=\left\langle \Delta v_{x}\Delta v_{y}\right\rangle _{LFK}^{\alpha/\beta}=\left\langle \Delta v_{x}\Delta v_{z}\right\rangle _{LFP}^{\alpha/\beta}\nonumber \\
 & =\left\langle \Delta v_{y}\Delta v_{z}\right\rangle _{LFP}^{\alpha/\beta}.
\end{align}
We observe that the numerical change of the first and second moments
approximately reproduce those of the Landau-Fokker-Planck operator
when two key conditions are met: a) $\tan^{2}\Theta/2\ll1$ (from
Eqs. (from Eqs. \ref{eq:<dvz>}-\ref{eq:<dvy2>}), which implies that
the angles sampled $\Theta$ from the normal distribution (Eq. \ref{eq:TA-f(tantheta)})
must be small (e.g., $\tan0.1\approx0.1$), and b) $\left\langle \tan^{4}\Theta/2\right\rangle \approx0$
(from Eq. \ref{eq:<dvz2>}), which, since $\left\langle \tan^{4}\Theta/2\right\rangle =3\sigma^{4}$,
it suggests that $\sigma^{2}\ll1$. It is worth noting that for species
$\alpha$ colliding with species $\beta$, the coefficients are proportional
to $n_{\beta}$, and vice-versa.

For a MC implementation, the crucial point is to \textcolor{red}{reproduce
the correct diffusion coefficients (otherwise the scheme would not
reproduce the correct Fokker-Plank equation)}, which necessitates
a suitable particle-paring scheme. The basic idea is then to first
pick a density for the variance of the normal distribution function
(Eq. \ref{eq:TA-f(tantheta)}). We then choose the number of particles
of each species to collide according to the collision probability.
The collision probability is chosen such that, after performing the
collisions, the rate of change of velocity moments on average should
approximately recover the Fokker-Planck coefficients, and it will
depend on the type of collisions and the species. We discuss the classical
same-weight TA treatment next, and extend to the variable-weight case
after.

\subsubsection{Classical TA particle-pairing scheme}

We first describe the classical TA method for pairing colliding particles.
Given two species $\alpha$ and $\beta$, each comprising $N_{\alpha}$
and $N_{\beta}$ particles within a cell, and corresponding densities
$n_{\alpha}=\gamma N_{\alpha}$ and $n_{\beta}=\gamma N_{\beta}$
, where $\gamma$ is a constant particle weight (with all particle
weights being equal for the time being), the process is as follows.

For \emph{self-collisions}, when there is an even number of particles,
pairs are randomly selected until all particles have collided. These
collisions employ the method previously described, with $n_{L}$ representing
the species density. In the case of an odd number of particles (assuming
three or more), three particles, denoted as $p_{1}$, $p_{2}$ and
$p_{3}$, are randomly selected. Three pairs are formed: $p_{1}$-$p_{2}$,
$p_{2}$-$p_{3}$, and $p_{3}$-$p_{1}$, each collision being conducted
via Eq. \ref{eq:TA-var}, with $n_{L}$ chosen as half the density
of the colliding species. Subsequently, the remaining even number
of particles are paired and collided accordingly.

For \emph{inter-species} particle pairing, assuming that $N_{\alpha}\leq N_{\beta}$
and $N_{\beta}=IN_{\alpha}+r$, where \textit{I} and $r<N_{\alpha}$
are positive integers, the process involves looping through the species-$\alpha$
particles $I$ times with each species-$\alpha$ particle pairing
with a species-$\beta$ particle via random sampling without replacement.
Following this, the remaining $r$ particles of species $\beta$ are
paired with species-$\alpha$ particles using the same random sampling
method. Collisions are conducted via Eq. \ref{eq:TA-var}, where $n_{L}$
is determined by the lower density of the colliding species.

The rationale that TA's pairing scheme effectively reproduces the
diffusion coefficients that account for the density of the colliding
species can be understood as follows. We take the inter-species collision
as an example,
\begin{align}
\left\langle \Delta v_{z}\right\rangle ^{\alpha/\beta} & =\frac{m_{\alpha\beta}}{m_{\alpha}}\left\langle \Delta u_{z}\right\rangle ^{\alpha},\nonumber \\
 & =-\frac{m_{\alpha\beta}}{m_{\alpha}\Delta t}u\left(1-\overline{\cos(\Theta)}\right)P_{c}^{\alpha}\nonumber \\
 & =-\frac{m_{\alpha\beta}}{m_{\alpha}\Delta t}u\overline{\left(\frac{2\tan^{2}\Theta/2}{1+\tan^{2}\Theta/2}\right)}\frac{N_{\beta}}{N_{\alpha}}\nonumber \\
 & \simeq-\frac{m_{\alpha\beta}}{m_{\alpha}}u\left(\frac{q_{\alpha}^{2}q_{\beta}^{2}n_{\alpha}\ln\Lambda}{4\pi\epsilon_{0}m_{\alpha\beta}^{2}u^{3}}\right)\frac{N_{\beta}}{N_{\alpha}}\nonumber \\
 & =-\frac{q_{\alpha}^{2}q_{\beta}^{2}n_{\beta}\ln\Lambda}{4\pi\epsilon_{0}m_{\alpha\beta}m_{\alpha}u^{2}},\label{eq:<Dvz>_TA}
\end{align}
which is the correct FP coefficient. Here $P_{c}^{\alpha}=\frac{N_{\beta}}{N_{\alpha}}=\frac{n_{\beta}}{n_{\alpha}}$
is the average probability of each collision, and the overbar denotes
averaged result over all possible scattering events, defined here
as $\overline{g(\delta)}=\int g(\delta)f(\delta)d\delta$ according
the distribution of Eq. \ref{eq:TA-f(tantheta)}. We have used $\overline{\tan^{2}\Theta/2}=\sigma^{2}$\textcolor{red}{{}
}, and $\sigma^{2}\ll1$ for the fourth step. As a result, we obtain
approximately the correct FP coefficient. It is apparent that the
ensemble-average of collisions involving species-$\alpha$ particles
colliding with species-$\beta$ ones yields coefficients proportional
to $n_{\beta}$, and vice-versa. For the special case of self-collisions,
most particles will collide once with average collision probability
unity, resulting in coefficients proportional to the species density.
In instances where the number of particles is odd, a few particles
collide twice with variance $\sigma^{2}/2$. This results in the correct
FP coefficients, proportional to the species density.

\subsubsection{Arbitrary particle-weight particle-pairing scheme}

The generalization of the TA pairing scheme to non-uniform weights
is relatively straightforward. Notably, our arbitrary-particle-weight
approach is different from Higginson's \citep{higginson2020corrected},
as we avoid employing repeated collisions for species of low-number
particles, and we incorporate corrections to ensure exact momentum
and energy conservation. 
\begin{algorithm}
\caption{\protect\label{alg:variable-weight=000020TA}Variable-weight TA algorithm
for a species $\alpha$ colliding with species $\beta$ within a cell.
In total, the algorithm will be performed over all cells and loop
over species. $N$ denotes the number of particles. We assume $n_{\alpha}\protect\leq n_{\beta}$,
where $n$ is obtained by summing \textcolor{red}{the in principle
unequal} particle weights \textcolor{red}{in a given cell }divided
by the cell volume.}

\begin{algorithmic}[1]
\footnotesize
\Ensure Shuffle all particles in species $\alpha, \beta$
\Procedure{Coulomb collisions}{$\alpha, \beta$}\Comment{Particle Coulomb collisions between species $\alpha$ and $\beta$}
\State $N_\alpha^c \gets N_\alpha$
\State $\tilde{N}_\beta^c \gets N_\beta n_\alpha/n_\beta$
\State $N_\beta^c \gets \mathrm{int}(\tilde{N}_\beta^c$)\Comment{int() is the integer part of the enclosed number}
\State $R_\beta^c \gets \tilde{N}_\beta^c-N_\beta^c$
\If{$R_\beta^c > 0$}
	\State $N_\beta^c = N_\beta^c+1$
\EndIf
\State $N_c = \mathrm{max}(N_\alpha^c,N_\beta^c)$\Comment{$N_\alpha \leq N_c\leq N_\beta$}
\State $i \gets 1$
\State $j \gets 1$
\State $k \gets 1$
\While{$k \leq N_c$} 
\State $\Delta\mathbf{u} \gets $ Eq.~\ref{eq:Delta-u-TA}
\If{$k \leq N_\alpha^c$}
	\State $\mathbf{v}_{i}^{p} \gets \mathbf{v}_{i} + (m_{\alpha\beta}/m_\alpha)\Delta\mathbf{u}$
\EndIf
\If{$k \leq N_\beta^c$}
	\If{$k==N_\beta^c$}
		\State $r \gets U(0,1)$ \Comment{$U(0,1)$ is a uniform random number between 0 and 1}
		\If{$r \geq R_\beta^c$}
			\State \textbf{continue} \Comment{no collision update for the particle $j$}
		\EndIf
	\EndIf
	\State $\mathbf{v}_{j}^{p} \gets \mathbf{v}_{j} - (m_{\alpha\beta}/m_\beta)\Delta\mathbf{u}$ 
\EndIf
\State $i \gets i+1$ 
\State $j \gets j+1$ 
\State $k \gets k+1$ 
\EndWhile\label{coll.pairs}
\EndProcedure 
\end{algorithmic}
\end{algorithm}
Suppose we have two species with non-uniformly weighted $N_{\alpha}$
and $N_{\beta}$ particles. Algorithm \ref{alg:variable-weight=000020TA}
details the extended particle pairing scheme between two species.
In our implementation, we randomly choose the number of species-$\beta$
particles without replacement (see the Algorithm for details), and
make them collide with randomly selected species-$\alpha$ particles.
Note that all $N_{\alpha}$ species-$\alpha$ particles collide once,
and only $N_{\beta}^{c}$ species-$\beta$ particles collide. Collisions
are always made pair-wise following the TA velocity update; however,
it may happen that only one particle updates its velocity during a
pair collision.

We show next that our variable-weight algorithm reproduces the FP
transport coefficients rigorously. Similarly to Ref. \citep{higginson2020corrected},
instead of taking the lower density in Eq. \ref{eq:TA-var}, we opt
for the higher density, $n_{H}$, such that 
\begin{equation}
\sigma^{2}=\frac{e_{\alpha}^{2}e_{\beta}^{2}n_{H}\ln\Lambda}{8\pi\epsilon_{0}m_{\alpha\beta}^{2}u^{3}}\Delta t.\label{eq:TA-var-h}
\end{equation}
We allow all of the lower-density particles to collide. Following
Eq. \ref{eq:<Dvz>_TA}, assuming $n_{\alpha}<n_{\beta}$ (so that
$n_{H}=n_{\beta})$, we extend the derivation of the FP coefficient
for the arbitrary particle-weight case as: 
\begin{align}
\left\langle \Delta v_{z}\right\rangle ^{\alpha/\beta} & =\frac{m_{\alpha\beta}}{m_{\alpha}}\left\langle \Delta u_{z}\right\rangle ^{\alpha},\nonumber \\
 & =-\frac{m_{\alpha\beta}}{m_{\alpha}\Delta t}u\left(1-\overline{\cos(\Theta)}\right)P_{c}^{\alpha}\nonumber \\
 & =-\frac{m_{\alpha\beta}}{m_{\alpha}\Delta t}u\overline{\left(\frac{2\tan^{2}\Theta/2}{1+\tan^{2}\Theta/2}\right)}\times1\nonumber \\
 & \simeq-\frac{m_{\alpha\beta}}{m_{\alpha}}u\left(\frac{e_{\alpha}^{2}e_{\beta}^{2}n_{\beta}\ln\Lambda}{4\pi\epsilon_{0}m_{\alpha\beta}^{2}u^{3}}\right)\nonumber \\
 & =-\frac{e_{\alpha}^{2}e_{\beta}^{2}n_{\beta}\ln\Lambda}{4\pi\epsilon_{0}m_{\alpha\beta}m_{\alpha}u^{2}},\label{eq:<Dvz>a_TA-e}
\end{align}
where we have used $P_{c}^{\alpha}=1$, as all species-$\alpha$ particles
collide exactly once. In contrast, for species $\beta$ we obtain:
\begin{align}
\left\langle \Delta v_{z}\right\rangle ^{\beta/\alpha} & =\frac{m_{\alpha\beta}}{m_{\beta}}\left\langle \Delta u_{z}\right\rangle ^{\beta},\nonumber \\
 & =-\frac{m_{\alpha\beta}}{m_{\beta}\Delta t}u\left(1-\overline{\cos(\Theta)}\right)P_{c}^{\beta}\nonumber \\
 & =-\frac{m_{\alpha\beta}}{m_{\beta}\Delta t}u\overline{\left(\frac{2\tan^{2}\Theta/2}{1+\tan^{2}\Theta/2}\right)}\times\frac{n_{\alpha}}{n_{\beta}}\nonumber \\
 & \simeq-\frac{m_{\alpha\beta}}{m_{\beta}}u\left(\frac{e_{\alpha}^{2}e_{\beta}^{2}n_{\alpha}\ln\Lambda}{4\pi\epsilon_{0}m_{\alpha\beta}^{2}u^{3}}\right)\nonumber \\
 & =-\frac{e_{\alpha}^{2}e_{\beta}^{2}n_{\beta}\ln\Lambda}{4\pi\epsilon_{0}m_{\alpha\beta}m_{\beta}u^{2}},\label{eq:<Dvz>a_TA-e-1}
\end{align}
where we have used $P_{c}^{\beta}=n_{\alpha}/n_{\beta}$ to achieve
the correct FP coefficient. Other FP coefficients can be derived similarly.

We note that, while the total momentum and energy are conserved on
average as the change in momentum and energy of each $\alpha$ particle
is counterbalanced by the average changes in the paired $\beta$ particles,
the total momentum and energy are not exactly conserved for any timestep.
To ensure exact momentum conservation, we implement a correction step
proposed by Tanaka et al. \citep{tanaka2018coulomb}:

\begin{equation}
\mathbf{v}_{i}^{p\prime}=\mathbf{V}_{0}+\alpha(\mathbf{v}_{j}^{p}-\mathbf{V}_{0}^{p}),\label{eq:post-coll-v-corr}
\end{equation}
where $\mathbf{V}_{0}=\sum_{i}m_{i}\mathbf{v}_{i}/\sum_{i}m_{i}$,
and the superscript p denotes post-collision quantities. The momentum
conservation is exactly satisfied by summing over all particles, i.e.,
$\sum_{i}m_{i}v_{i}^{p\prime}=\sum_{i}m_{i}V_{0}$. The correction
factor $\alpha$ is found by enforcing the energy conservation:
\begin{equation}
\sum_{i}w_{i}m_{i}\left(v_{i}^{p\prime}\right)^{2}/2=E_{tot},\label{eq:correct-energy}
\end{equation}
where $E_{tot}$ is the total energy before collisions. By substituting
$v_{i}^{p\prime}$ into Eq. \ref{eq:post-coll-v-corr}, we find that:
\begin{equation}
\alpha=\frac{E_{tot}-0.5\sum_{i}m_{i}(V_{0})^{2}}{E_{tot}^{p}-0.5\sum_{i}m_{i}(V_{0}^{p})^{2}}.
\end{equation}

\subsection{Collisional coupling with a fluid-electron component in a hybrid
kinetic-ion/fluid-electron model}

In many instances, hybrid kinetic-ion/fluid-electron formulations are of interest (and in fact will be used in one of our numerical tests). These formulations neglect electron kinetic-scale effects and model the electrons as a simplified fluid that is discretized on a spatial mesh~\cite{stanier2019,winske2023hybrid}. The electric field is found from the electron momentum equation (Ohm's law), where electron inertia is typically neglected and a scalar electron pressure ($p_e$) is used. For instance, in the electrostatic limit, the Ohm's law reads:
\begin{equation}\label{ohms}\boldsymbol{E} = -\frac{\boldsymbol{\nabla} p_e - \boldsymbol{S}_{\boldsymbol{P}}}{en_e}.\end{equation}
Here, $n_e = \tfrac{1}{e}\sum_\alpha Z_\alpha n_\alpha$ is the quasineutral electron density, and $\boldsymbol{S}_{\boldsymbol{P}}$ is the momentum source due to collisions with the kinetic ion particles (defined below). The scalar electron pressure $p_e = n_e k T_e$, with $T_e$ the electron temperature, is evolved from:
\begin{equation}\label{pressure}\partial_t p_e + \gamma_h \boldsymbol{\nabla} \cdot \left(\boldsymbol{u}_e p_e \right) + \left(\gamma_h - 1\right)\left(\boldsymbol{\nabla} \cdot \boldsymbol{q}_e - \boldsymbol{u}_e \cdot \boldsymbol{\nabla} p_e \right)=0.\end{equation}
Here, $\boldsymbol{u}_e = \tfrac{1}{en_e}\sum_\alpha Z_\alpha n_\alpha \boldsymbol{u}_\alpha$ is the ambipolar electron bulk velocity, $\gamma_h=5/3$ is the ratio of specific heats.  

 \textcolor{red}{We note that the energy change in electrons due to collisions with ions has already been taken into account by evolving the temperature of the electron Gaussian during in the collisional step but, since the ambipolarity condition above used to set $\boldsymbol{u}_e$ overwrites any change to the electron gaussian velocity, the momentum source $\boldsymbol{S}_{\boldsymbol{P}}$ is still required in Eq.~(\ref{ohms}) to account for collisional friction.} The electron momentum source $\boldsymbol{S}_{P}$ due to collisions with kinetic ions are computed as~\cite{sherlock2008monte}:
\begin{equation}\boldsymbol{S}_{P} = -\frac{\sum_p m_p \left(\mathbf{v}_{p}^\prime - \mathbf{v}_{p}\right)}{\Delta t \Delta V},\end{equation}
where the sum is performed over all particles within a given cell of volume $\Delta V$, $\boldsymbol{v}_p$ are the pre-collision velocities, and $\boldsymbol{v}_p^\prime$ are the post-collision velocities. As this source is directly calculated from the total ion momentum change, using it in Eq.~(\ref{ohms}) ensures total conservation of momentum.   

The heat flux $\boldsymbol{q}_e$ and the particle source terms are dependent on the choice of closure for the electron fluid equations. We use the collisional Braginskii-like~\cite{braginskii1958transport} closure of Ref.~\cite{simakov2014electron} that is valid for multiple species of ions, and has good accuracy for arbitrary values of the parameter:
\begin{equation}Z_\textrm{eff} = \frac{\sum_\alpha Z_\alpha^2 n_\alpha}{en_e}.\end{equation}
The electron heat flux is calculated as ~\cite{simakov2014electron}:
\begin{equation}\label{heatflux}\boldsymbol{q}_e = \beta_0 p_e \left(\boldsymbol{u}_e -  \langle \boldsymbol{u}_\alpha \rangle\right)  - \kappa \boldsymbol{\nabla} T_e,\end{equation}
where $\langle \boldsymbol{u}_\alpha \rangle$ is the collision-frequency averaged velocity, given by:
\begin{equation}\langle \boldsymbol{u}_\alpha \rangle = \frac{\sum_\alpha \nu_{e\alpha} \boldsymbol{u}_{\alpha}}{\sum_\alpha \nu_{e\alpha} },\end{equation}
and 
\begin{equation}\nu_{e\alpha} = \frac{n_\alpha Z_\alpha^2 e^4 \ln{\Lambda}}{6\sqrt{2}\pi^{3/2}\left(kT_e\right)^{3/2} \sqrt{m_e}\epsilon_0^2}.\end{equation}
The heat conductivity is given by:
\begin{equation}\kappa = \frac{\gamma_0 p_e}{m_e \sum_\alpha \nu_{e\alpha}},\end{equation}
where the coefficients $\beta_0$, $\gamma_0$ are defined~\cite{simakov2014electron} as
\begin{equation}\beta_0 = \frac{30Z_\textrm{eff}\left(11Z_\textrm{eff}+15\sqrt{2}\right)}{217Z_\textrm{eff}^2 + 604\sqrt{2}Z_\textrm{eff} + 288},\end{equation}
\begin{equation}\gamma_0 = \frac{25Z_\textrm{eff}\left(433Z_\textrm{eff}+180\sqrt{2}\right)}{4\left(217Z_\textrm{eff}^2 + 604\sqrt{2}Z_\textrm{eff} + 288\right)}.\end{equation}

Section \ref{subsec:lemons} above describes the methods used to collide the ion species with the electron fluid, where the ions can be treated as either particles or Gaussians, depending on their collisionality. As described, the methods can account for the ion species colliding with a bulk Maxwellian electron fluid with a given density $n_e$, temperature $T_e$, and drift velocity $\boldsymbol{u}_e$. However, we must also include collisional effects in the ion species due to the non-Maxwellian part of the electron distribution function that was already used in the derivation of Eq.~\ref{heatflux}. This correction is done here following a similar approach to that used in Ref.~\cite{sherlock2008monte,le2023hybrid}. Firstly, the collisional friction between particle ions and electrons is calculated using the closure of Ref.~\cite{simakov2014electron} as:
\begin{equation}\boldsymbol{F}_{\alpha e} = -m_e n_e \nu_{e \alpha} \left(\boldsymbol{u}_\alpha - \langle \boldsymbol{u}_\alpha \rangle \right)+ \alpha_0 m_e n_e \nu_{e \alpha} \left(\boldsymbol{u}_e -  \langle \boldsymbol{u}_\alpha \rangle \right) + \beta_0 \frac{n_e \nu_{e \alpha} \boldsymbol{\nabla} T_e}{\sum_\beta \nu_{e \beta}},\end{equation}
where
\begin{equation}\alpha_0 = \frac{4\left(16Z_\textrm{eff}^2 + 61\sqrt{2}Z_\textrm{eff} + 72\right)}{217Z_\textrm{eff}^2 + 604\sqrt{2}Z_\textrm{eff} + 288}.\end{equation}
This expression contains both the bulk-Maxwellian contribution ($f_0$) as well as the $\delta f$ contributions. To avoid double counting the bulk-ion/Maxwellian-electron contribution, which is already taken care of in the standard collisional step in Sec.~\ref{subsec:lemons}, the Maxwellian-electron contribution is subtracted and the remainder friction force is cast into a species independent form (by dividing by $Z_\alpha^2e^2n_\alpha$) as
\begin{equation}\boldsymbol{\delta F} = \frac{\boldsymbol{F}_{\alpha e} - m_e n_e \nu_{e\alpha} \left(\boldsymbol{u}_e - \boldsymbol{u}_\alpha\right)}{Z_\alpha^2 e^2n_\alpha} = \frac{\nu_{e\alpha}}{Z_\alpha^2e^2 n_\alpha}\left[\left(\alpha_0-1\right) m_en_e \left(\boldsymbol{u}_e - \langle \boldsymbol{u}_\alpha \rangle\right) + \beta_0 n_e \frac{\boldsymbol{\nabla}T_e}{\sum_\beta \nu_{e\beta}}\right].\end{equation}
This expression, which is computed on the mesh, can then be used to correct the post-collision ion-particle velocities as:
\begin{equation}\frac{d \boldsymbol{v}_p}{dt} = \frac{q_p^2}{m_p}\boldsymbol{\delta F} \left(\boldsymbol{x}_g\right)S_0 \left(\boldsymbol{x}_g - \boldsymbol{x}_p\right),\end{equation}
for particle $p$ at position $\boldsymbol{x}_p$ with charge $q_p$ and mass $m_p$. Here $\boldsymbol{x}_g$ is the position of the cell centers and $S_0$ is the nearest-grid-point (top hat) interpolation function.

\subsection{Hybrid Maxwellian-MC algorithm orchestration}

\begin{figure}
\begin{centering}
\includegraphics[width=1\columnwidth]{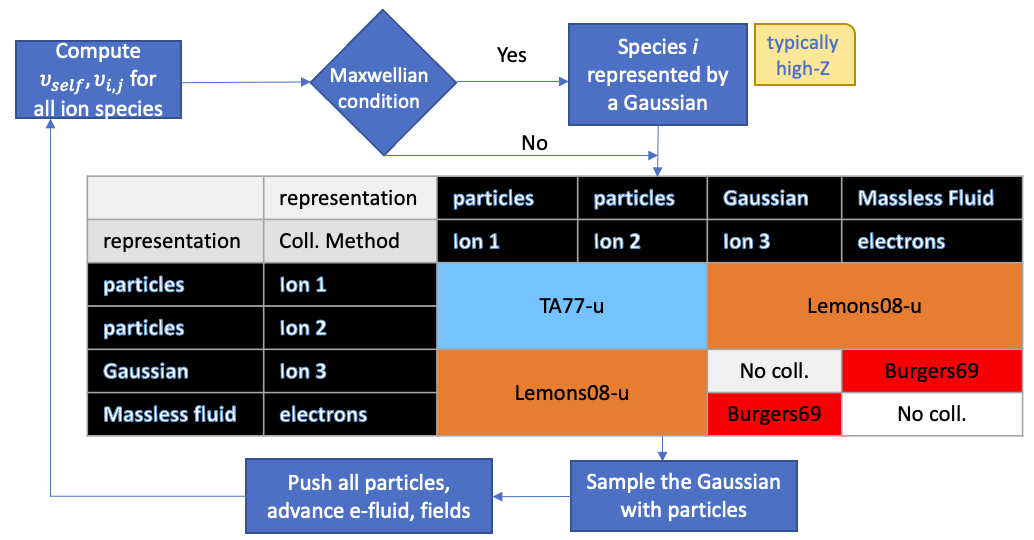}
\par\end{centering}
\caption{\protect\label{Fig:HMMC-algorithm}Flowchart of an HMMC step coupled
with other parts of the hybrid model. For the purposes of illustration,
the flowchart assumes two kinetic ion species, a Maxwellian ion species,
and a massless electron fluid species. \textcolor{red}{In the collision
table, ``TA77-u'' refers to the upgraded TA method (with variable
particle weights), ``Lemons08-u'' refers to the upgraded Lemons
method (with low-relative-speed treatment), and ``Burgers69'' refers
to the Burgers method.}}
\end{figure}

Figure \ref{Fig:HMMC-algorithm} illustrates the orchestration of
the hybrid particle-Maxwellian collisions algorithm, which comprises
three types of collisions: Maxwellian-Maxwellian, particle-Maxwellian,
and particle-particle. This algorithm can deal with any plasma collisionality,
but is particularly advantageous for plasma systems featuring a combination
of weakly and strongly collisional species, where one or more species
can be effectively represented by Maxwellian distribution functions.
The criterion for selecting the Maxwellian approximation is the magnitude
of the self-collision frequency becoming stiff, i.e. $\nu_{self}\Delta t>1$,
except when the number of particles in that cell is too low for a
Maxwellian reconstruction (we typically use four particles as the
threshold), in which case we revert back to TA.

The algorithm functions as follows: it iterates through species pairs
for collision events. For each pair, the appropriate collision model
is selected from the three considered in this study. If a species
is deemed to follow a Maxwellian distribution, the advancement of
its first five moments due to collisions is determined. Upon completion
of all collision pairs, an additional step samples particles from
the post-collision Maxwellian distributions.\textcolor{red}{{} Following
this random sampling, a shift-and-scale adjustment is applied to ensure
exact conservation of both momentum and energy} \citep{chen2021unsupervised}.
These particle samples are subsequently utilized for the Vlasov transport
step.

As described in the previous section, conservation of mass, momentum,
and energy are strictly enforced for each collisional process considered
in the algorithm. As a result, the total mass, momentum, and energy
conservation is conserved exactly (in practice, to numerical round-off
error) after all collisional processes have been performed.

\section{Numerical experiments}

\label{sec:Numerical-experiments}

We have implemented the HMMC algorithm in C++, leveraging the parallelization
strategies offered by the Cabana \citep{mniszewski2021enabling} and
Kokkos \citep{edwards2014kokkos} libraries. We conduct a comprehensive
evaluation of our proposed HMMC method and its components with a series
of numerical tests intended to highlight its advantages, capabilities,
and potential advancements over existing approaches. Our numerical
experiments start with a two-species relaxation test intended to validate
the extended TA model with non-equal weights introduced earlier in
this study (see Sec. \ref{subsec:extended-TA}).

We continue with particle-Maxwellian collisions, where we compare
the standard Lemons method \citep{lemons2009small} with the enhanced
one proposed in this study, with a focus on particles exhibiting low
relative and thermal speeds. Given their prolonged collision times
with the Maxwellian distribution, these low-relative speed particles
are expected to undergo significant scattering angles, leading to
isotropic scattering. This behavior necessitates a distinct modeling
approach from that for particles with higher relative speed.

Subsequently, we test the full hybrid method (i.e., considering all
possible collisional processes) with a challenging relaxation problem
involving He-C-Au-e interactions. This scenario, which is inspired
by conditions usually encountered in ICF hohlraums, includes realistic
high-mass-ratios and high-Z species in a stiff and self-consistently
coupled system relaxing collisionally.

Finally, we consider the coupling of our HMMC approach with a fully
self-consistent PIC-ion/fluid-electron model to simulate a 1D plasma
interpenetration problem in typical ICF hohlraum conditions, inspired
by recent experiments conducted by Le Pape et al.~\citep{LePape2020},
demonstrating the ability of the HMMC method to capture transport
in realistic plasma scenarios.

\subsection{Overview of the iFP Vlasov-Fokker-Planck Code}

For verification of the HMMC algorithm, we compare our results against
simulations obtained with the plasma kinetic code iFP \citep{taitano2018adaptive,Taitano2021,Taitano2021b}.
iFP is an Eulerian Vlasov-Fokker-Planck code that solves for the ion
distribution functions on a phase-space grid. The iFP code is one-dimensional
in physical space and two-dimensional in velocity-space ($v_{\parallel},v_{\bot}$),
a valid configuration under a 1D electrostatic approximation without
loss of generality. Electrons are modeled as quasineutral and ambipolar
fluid with zero charge and current density, possessing their own temperature.
iFP solves for the fully Landau-Fokker-Planck collisional model, using
the Rosenbluth formulation for optimal performance. To further reduce
computational cost, iFP employs mesh adaptivity in the physical space
through a mesh-motion scheme that optimizes the grid to resolve features
such as gradients of the moment quantities. Additionally, iFP utilizes
a mesh-transformation strategy in velocity-space, where each species'
velocity-space mesh is shifted based on its bulk velocity and scaled
according to its thermal speed. iFP has been designed to conserve
mass, momentum, and energy exactly (in practice, to nonlinear tolerance)
for both Vlasov and collision-operator components, making it uniquely
suited for verification purposes. In the tests that follow, unless
otherwise specified, iFP employs a velocity-space grid resolution
$[N_{v_{\parallel}},N_{v_{\bot}}]=[256,128]$, with domain extent
{[}$v_{\parallel,\min},v_{\parallel,\max}]=[-7v_{th},7v_{th}]$, {[}$v_{\bot,\min},v_{\bot,\max}]=[0,7v_{th}]$.

\textcolor{red}{To facilitate comparison with iFP for the relaxation
tests, we normalize time scales with respect to the proton-proton
collision frequency \citep{hinton1983collisional,Taitano2021} at
a reference state $n^{*}$, $T^{*}$ (with reference Coulomb logarithm
$\ln\Lambda^{*}=10$ unless otherwise stated),
\begin{equation}
\nu_{pp0}=n^{*}e{}^{4}\textrm{ln}\Lambda^{*}/12\pi^{3/2}\epsilon_{0}^{2}m_{p}^{1/2}(kT^{*})^{3/2}.\label{eq:nu_pp}
\end{equation}
If we normalize $\hat{n}=n/n^{*}=1$, $\hat{kT}=kT/kT^{*}=1$, and
$\hat{\ln\Lambda}=\ln\Lambda/\ln\Lambda^{*}=1$, then $\hat{\nu}_{pp}=\nu_{pp}/\nu_{pp0}=1$
for proton-proton collisions.}

\subsection{Unequal-weight MC algorithm}

\begin{figure}
\begin{centering}
\includegraphics[width=0.5\columnwidth]{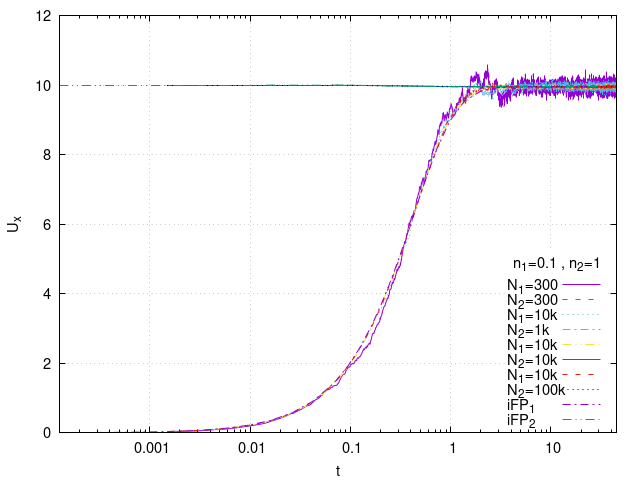}\includegraphics[width=0.5\columnwidth]{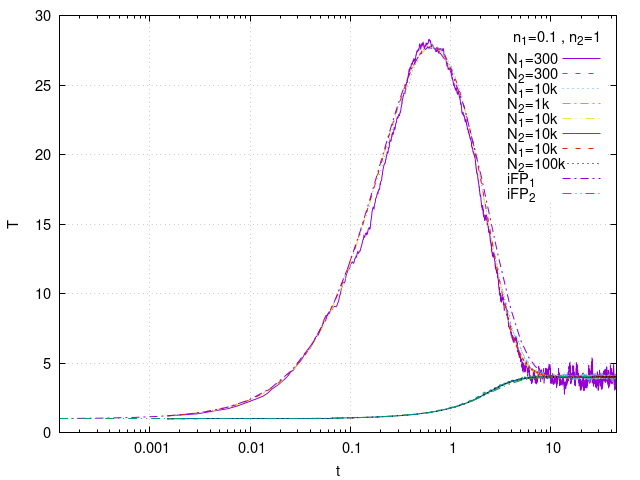}
\par\end{centering}
\caption{\protect\label{Fig:2sp-relaxation-noneqw}Comparison of the two-species
relaxation test results using the TA algorithm with non-equal particle
weights for various particle-number choices and with iFP. All simulations
are in excellent agreement.}
\end{figure}

\begin{figure}
\begin{centering}
\includegraphics[width=0.5\columnwidth]{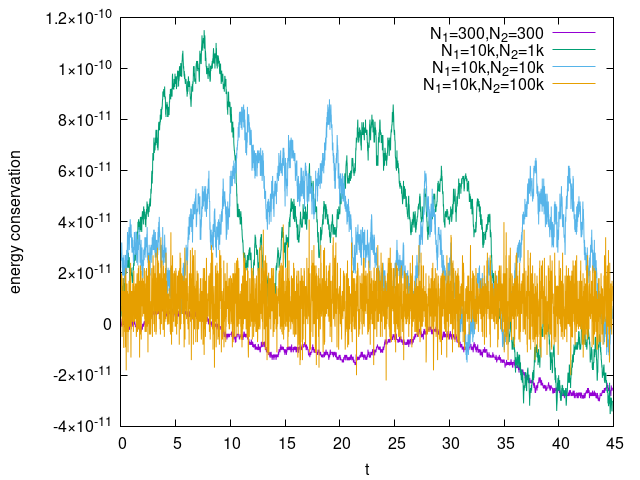}\includegraphics[width=0.5\columnwidth]{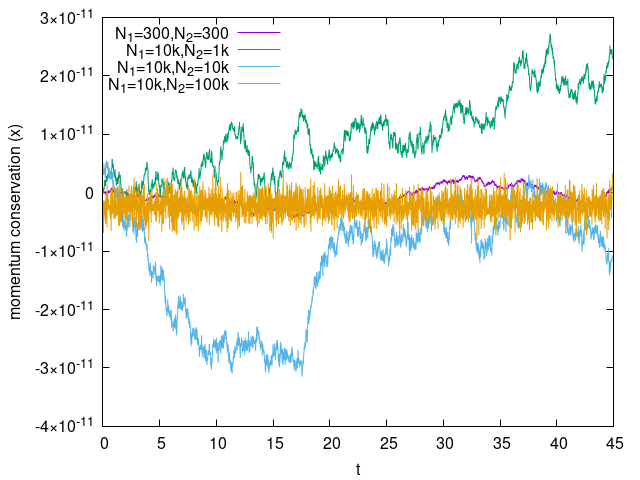}
\par\end{centering}
\begin{centering}
\includegraphics[width=0.5\columnwidth]{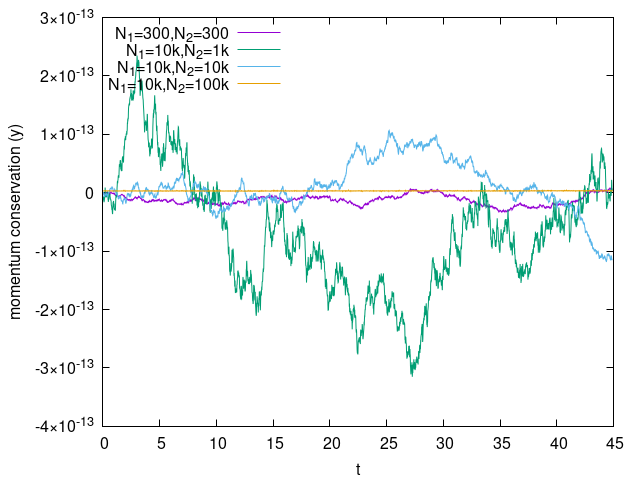}\includegraphics[width=0.5\columnwidth]{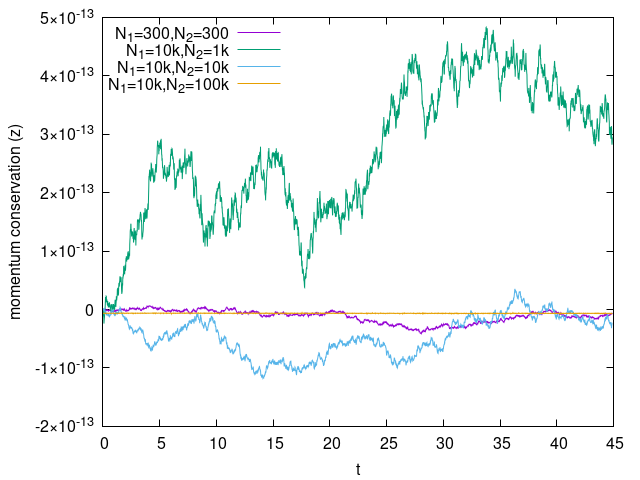}
\par\end{centering}
\caption{\protect\label{fig:Conservation-noneqw-TA}Conservation properties
of non-equal weight TA method. Errors (relative to the initial values)
of conservation of energy (top-left) and x-momentum (top-right) are
at the $10^{-11}$ level. Errors in the other two momenta (y and z
directions, which have zero initial momentum) are conserved to $<10^{-13}$.}
\end{figure}

To assess the unequal-particle-weight TA method, we follow Ref. \citep{higginson2020corrected}
and consider a two-species relaxation problem with initial conditions
outlined in Table \ref{Table:2-sp-noneqw}.
\begin{table}
\centering{}\caption{\protect\label{Table:2-sp-noneqw}Initial conditions for the two-species
relaxation test. The problem features a density ratio of 10.}
\begin{tabular}{|c|c|c|}
\hline 
 & species 1 & species 2\tabularnewline
\hline 
\hline 
mass & 1 & 20.0\tabularnewline
\hline 
Z number & 1 & 20.0\tabularnewline
\hline 
density & 0.1 & 1.0\tabularnewline
\hline 
drift velocity & 0.0 & 10.0\tabularnewline
\hline 
thermal velocity (x, y, z) & 1.0 & 0.2236\tabularnewline
\hline 
\end{tabular}
\end{table}
 The results are presented in Figure \ref{Fig:2sp-relaxation-noneqw}.
Four runs with varying numbers of particles are conducted: $N_{1}=N_{2}=300$,
$N_{1}=10^{4}$ and $N_{2}=10^{3}$, $N_{1}=N_{2}=10^{4}$, and $N_{1}=10^{4}$
and $N_{2}=10^{5}$. Given a density ratio between the two species
$n_{2}/n_{1}=10$, the weight ratios for the four particle-number
configurations are $w_{2}/w_{1}=10,100,10,1$, respectively. The run
with $N_{1}=N_{2}=300$ is the noisiest, as expected. However, there
is excellent agreement between runs with different particle weights.\textcolor{red}{{}
}Figure \ref{fig:Conservation-noneqw-TA} shows the history of the
conservation properties of the various particle weighting numerical
experiments. For all cases, the noise level of the conservation of
energy and momentum is at or better than $10^{-.10}$, improving with
the number of sample particles, as expected. If the conservation error.
is normalized to the number of particles, the average error is $\lesssim10^{-14}$,.
of order the double-precision round-off error.

\subsection{Improved particle-Maxwellian (fluid) Lemons algorithm}

In Ref. \citep{lemons2009small}, the Lemons method (also known as
``particle-moment'' method) was tested with two test problems: a
single-component plasma, and two equal density, equal mass Maxwellian
components with different initial temperatures. In the reference,
they were able to use a timestep about 10\% of the relaxation time,
which is much larger than that allowed by the classic TA method. Such
performance gain was attributed to the fact that the method deals
with collisions of test particles with the target Maxwellian distribution
instead of colliding individual particles. However, the failure mode
of the original method identified in Sec. \ref{subsec:lemons} cannot
be detected with the two tests used in the reference.

\begin{figure}
\begin{centering}
\includegraphics[width=0.5\columnwidth]{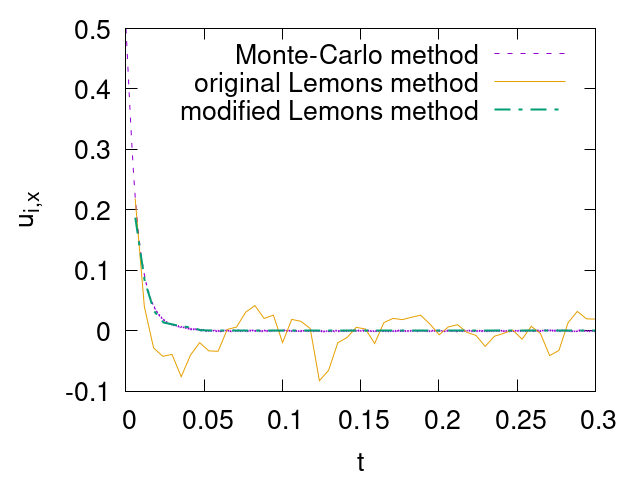}\includegraphics[width=0.5\columnwidth]{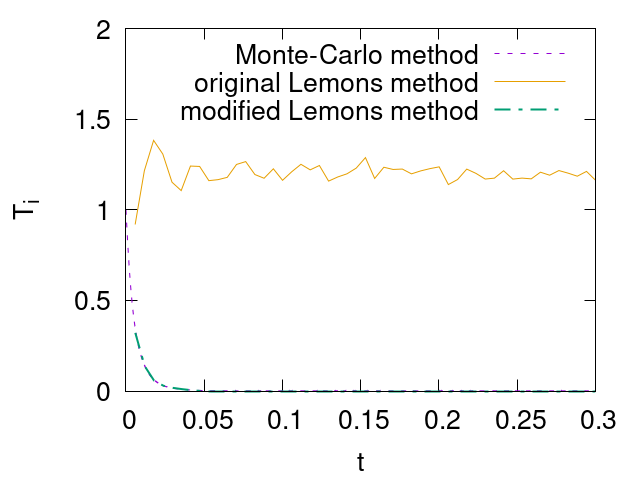}
\par\end{centering}
\caption{\protect\label{Lemons-UT-relax}Comparison of \textcolor{red}{ion-}species
momentum and temperature relaxation simulations using Lemons, improved
Lemons, and TA methods with simulation parameters as described in
the main text. Incorrect results are obtained with the original Lemons
method\textcolor{red}{. }The improved Lemons method produces correct
results, as assessed by comparison with a temporally resolved TA simulation
using a 4000$\times$ smaller timestep.}
\end{figure}

Here, we compare the original Lemons method, our proposed improved
version, and the classic TA method using a relaxation problem of a
homogeneous particle-ion/fluid-electron plasma (i.e., with disparate
masses). In the test, for the ion species we take $m_{i}=1$, $q_{i}=1$,
$u_{i0}=0.5$, and $T_{i0}=1$. The electron mass is 0.01, the electron
charge is $q_{e}=-1$, and the electron \textcolor{red}{velocity and
}temperature \textcolor{red}{are $u_{e0}=0$ and} $T_{e0}={\color{red}1.836\times10^{-3}}$\textcolor{red}{,
which are kept fixed during the simulation}. Both species are initialized
as Maxwellians. We employ $N_{p}=1000$, ${\color{red}\Delta t_{L}=6\times10^{-3}\nu_{pp0}^{-1}}$
for Lemons, and $N_{p}=10,000$, ${\color{red}\Delta t_{TA}}\mathrel{\color{red}=}{\color{red}1.5\times10^{-6}\nu_{pp0}^{-1}}$
for TA \textcolor{red}{(note $\nu_{ie}\Delta t_{L}\sim0.6$, which
is significant, while $\nu_{TA}\Delta t_{TA}\sim0.015$ for accuracy,
and that $\nu_{TA}/\nu_{ie}\propto m_{i}/m_{e}=100$, so the Lemons
timestep is much larger than the TA one by about a factor of 4000)}.
The results are shown in Fig. \ref{Lemons-UT-relax}. Clearly, the
original Lemons method fails to capture the right behavior of the
temperature relaxation, and features significant fluctuations due
to enhanced particle noise. The improved Lemons method not only agrees
very well with TA, but uses 10 times fewer particles and \textcolor{red}{4000$\times$}
larger timesteps. The overall efficiency gain of Lemons vs. TA for
this case is $\mathrel{\color{red}>}{\color{red}10^{4}}$.\textcolor{red}{{}
It is also worth noting that the performance impact of the low-relative-speed
treatment is negligible (<1\% CPU time of the Lemons collision kernel).}

\subsection{Full HMMC algorithm (0D)}

We verify next the full HMMC algorithm self-consistently with a 0D
multi-species system comprising four species: Helium, Carbon, Gold,
and electrons. Simulation initial conditions are as described in Table
\ref{Table:4-sp-relaxation}. In this setup, the electron and Gold
species are described by a fluid model, while the rest are treated
as particles. The simulation incorporates all collisional interactions
among species.

In this system, Gold self-collisions exhibit the fastest timescales,
owing to its high atomic number ($\nu\propto Z^{4}$). Specifically,
the Gold self-collision frequency is approximately one to two orders
of magnitude faster than any other collision timescales in the system.
This poses a stiffness challenge if resolved temporally in the simulation,
as required by the TA method for accuracy \citep{cohen2010time}.
To avoid this, we represent the Gold species as a Maxwellian distribution,
while the other two ion species are treated with particles.

\begin{figure}
\begin{centering}
\includegraphics[width=0.5\columnwidth]{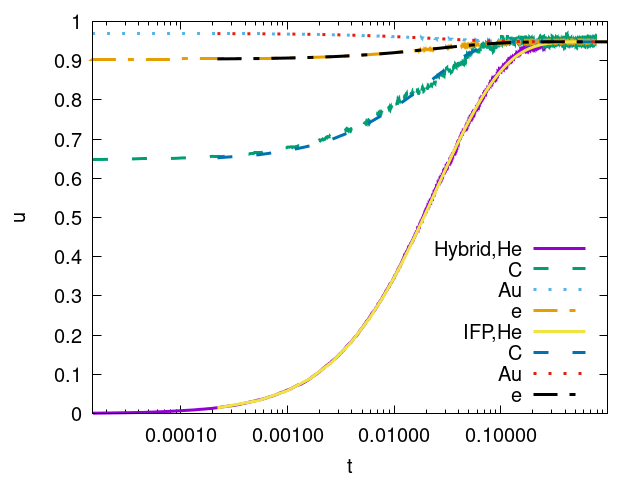}\includegraphics[width=0.5\columnwidth]{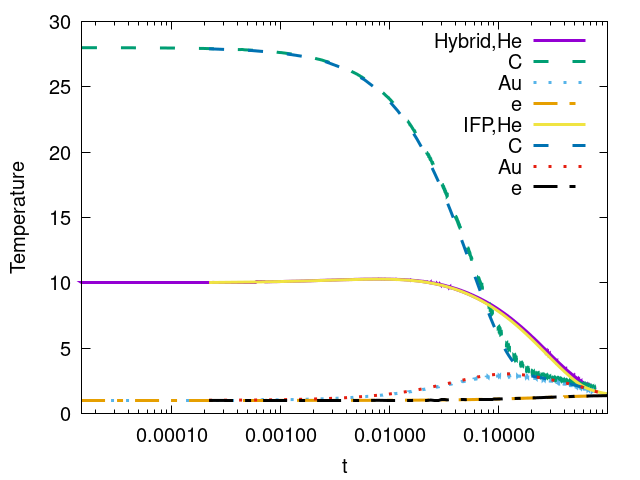}
\par\end{centering}
\caption{\protect\label{Fig:HeCAue-relaxation}Comparison of hybrid collision
algorithm with the iFP for a four-species relaxation problem. Here.
the Gold species is represented by a drift Maxwellian distribution.
The number of particles used for the kinetic species He and C are
$10^{5}$ and $10^{4}$, respectively.}
\end{figure}

\begin{figure}
\begin{centering}
\includegraphics[width=0.5\columnwidth]{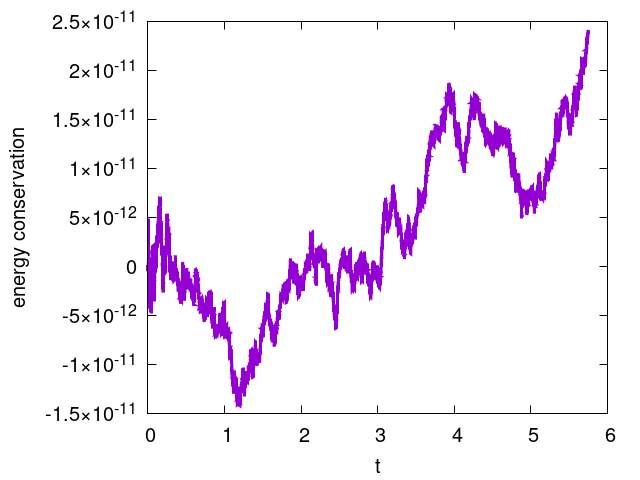}\includegraphics[width=0.5\columnwidth]{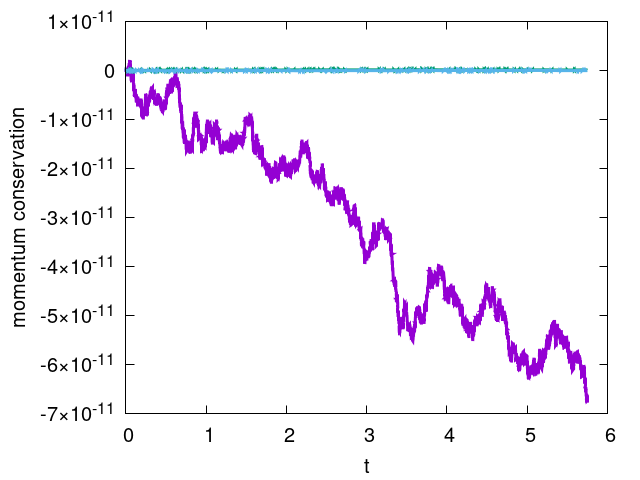}
\par\end{centering}
\caption{\protect\label{Fig:HeCAue-conservation}Conservation properties of
HMMC algorithm. Errors of conservation of energy (left) and the $x$-component
of momentum (right, magenta) is realized to to $10^{-11}$ level.
The y and z components of momentum (cyan) are conserved to $10^{-14}$
level.}
\end{figure}

Figure \ref{Fig:HeCAue-relaxation} presents the results of momentum
and energy relaxation for the four species from the hybrid model,
demonstrating very good agreement with iFP. Figure \ref{Fig:HeCAue-conservation}
demonstrates strict conservation of momentum and energy of HMMC to
near double precision when the error is normalized by the number of
particles. In terms of efficiency, for this simulation HMMC is about
\textcolor{red}{112}$\times$ faster than TA when the latter resolves
the fastest collisional time scale\textcolor{red}{, as needed for
accuracy \citep{cohen2010time} (the speedup calculation follows from
HMMC using $\nu_{Au}\Delta t=1.12$, where $\nu_{Au}$ is the frequency
of Au self-collisions, and assuming $\nu_{Au}\Delta t_{TA}=0.01$)}.

\begin{table}
\centering{}\caption{\protect\label{Table:4-sp-relaxation}Initial condition for the 4-species
relaxation. The problem is set to have a density ratio of 10.}
\begin{tabular}{|c|c|c|c|c|}
\hline 
 & Helium & Carbon & Gold & e-fluid\tabularnewline
\hline 
\hline 
mass & 4 & 12 & 197 & 1/1837\tabularnewline
\hline 
Z number & 2 & 6 & 30 & 1\tabularnewline
\hline 
density & 1.0 & 0.1 & 1.0 & 32.6\tabularnewline
\hline 
drift velocity & 0.0 & 0.6462 & 0.9693 & 0.9329\tabularnewline
\hline 
thermal velocity (x,y,z) & 1.5811 & 1.5275 & 0.07125 & 42.8602\tabularnewline
\hline 
\end{tabular}
\end{table}
.

\subsection{A 1D multispecies plasma interpenetration problem in a hohlraum-like
environment}

Our final demonstration tests the collective behavior of multi-species
relaxation and transport in a plasma interpenetration problem inspired
by the recent experiments of Le Pape et. al. \citep{LePape2020}.
In this experiment, a high-density carbon puck (density $3.45\mathrm{g/cc}$,
radius $600\mu\mathrm{m}$) is centered within a circular gold band
(density $19.3\mathrm{g/cc}$, inner radius $1600\mu\mathrm{m}$,
thickness $25\mu\mathrm{m}$), with a low-density helium gas fill
($0.15\mathrm{mg/cc}$) present between. The carbon and inner gold
surfaces are subjected to a laser-driven ablation via $351\mathrm{nm}$
laser light, which is intended to simulate the plasma environment
within an indirect-drive inertial confinement fusion (ICF) hohlraum.
The carbon and gold produce coronal plasma blowoff which counterpropagates
and collides in the middle of the domain, compressing the helium fill.
This is a very stringent numerical test due to the scale separation
present, and the sensitivity of the species interpenetration and temperature
transport to the accuracy of the collisional process.

Here, we model a planar-geometry surrogate of the Le Pape experiment
with static boundary conditions (non-moving, non-time-dependent),
which is intended to replicate the experimental plasma conditions.
The initial and boundary conditions are chosen to mimic the plasma
state in the coronal blowoff below the critical density, soon after
the laser drive starts. We choose representative average ionization
states for the ion species to be $Z_{He}=2.0$, $Z_{C}=6.0$, $Z_{Au}=32.0$.
The boundaries are located at $0\mu\mathrm{m}$ (carbon) and $1000\mu\mathrm{m}$
(gold). The initial/boundary parameters are chosen for carbon, gold,
and helium to be $n_{C,0}=2\times10{}^{26}\mathrm{m}^{-3}$, $T_{C,0}=2\mathrm{keV}$,
$u_{\parallel,C,0}=250\mathrm{km/s}$, $n_{Au,0}=2\times10{}^{26}\mathrm{m}^{-3}$,
$T_{Au,0}=0.75\mathrm{keV}$, $u_{\parallel,Au,0}=-1.5\mathrm{km/s}$,
$n_{He,0}=2.25\times10{}^{25}\mathrm{m}^{-3}$, $T_{He,0}=26.83\mathrm{eV}$,
and $u_{\parallel,He,0}=0\mathrm{km/s}$. The carbon and gold number
density initial spatial profiles are computed using a hyperbolic cosine
switch,
\begin{equation}
n_{C,Au}(x)=n_{C,Au,0}\frac{1}{\left[\cosh\left(\frac{x-x_{C,Au,0}}{\lambda}\right)\right]^{2}},
\end{equation}
with $\lambda=62.5\mu\mathrm{m}$, $x_{C,0}=0\mu\mathrm{m}$ is left
boundary for $C$ and $x_{Au,0}=1000\mu\mathrm{m}$ right boundary
for $Au$. Note that for the iFP simulations, zero particle density
is not possible, so a number density floor of $4.2\times10^{20}\,\mathrm{m}^{-3}$
is used to approximate near-vacuum conditions. The initial spatial
profile of temperature for carbon and gold and density for helium
taken to be uniform (equal to the parameters $T_{C,0}$, $T_{Au,0}$,
$n_{He,0}$, respectively). For a collisionally quiescent initial
condition, the initial spatial temperature profiles of the helium
and the electrons are set to be:
\begin{equation}
T_{e,He}(x)=\frac{T_{He,0}Z_{He}n_{He}+T_{C,0}Z_{C}n_{C}+T_{Au,0}Z_{Au}n_{Au}}{Z_{He}n_{He}+Z_{C}n_{C}+Z_{Au}n_{Au}},
\end{equation}
and the initial bulk velocity profiles of all ion species are computed
as:
\begin{equation}
u_{\parallel,C,Au,He}(x)=\frac{u_{\parallel,He,0}n_{He}+u_{\parallel,C,0}n_{C}+u_{\parallel,Au,0}n_{Au}}{n_{He}+n_{C}+n_{Au}}.
\end{equation}
The electrons are solved as a quasineutral/ambipolar fluid, thus their
number density and bulk velocity are determined at all times from
the conditions:
\begin{equation}
\begin{aligned}n_{e} & =\sum_{\alpha\in[C,Au,He]}Z_{\alpha}n_{\alpha},\\
u_{\parallel,e} & =\frac{\sum_{\alpha\in[C,Au,He]}Z_{\alpha}n_{\alpha}u_{\parallel,\alpha}}{n_{e}}.
\end{aligned}
\end{equation}

Our HMMC collisional algorithm has been implemented in a 1D hybrid
particle-in-cell code solving the exact same set of equations as iFP
\citep{taitano2018adaptive}. The hybrid particle in cell code is
operator-split, performing a collision timestep first (using HMMC),
followed by a standard leap-frog field/particle advance. The average
number of particles per cell is 1070, 594, and 594 for the He, C,
and Au species, respectively. We employ 64 uniform cells and $\Delta t=0.00234$
ps in the hybrid simulation.

\begin{figure}
\begin{centering}
\includegraphics[width=0.33\columnwidth]{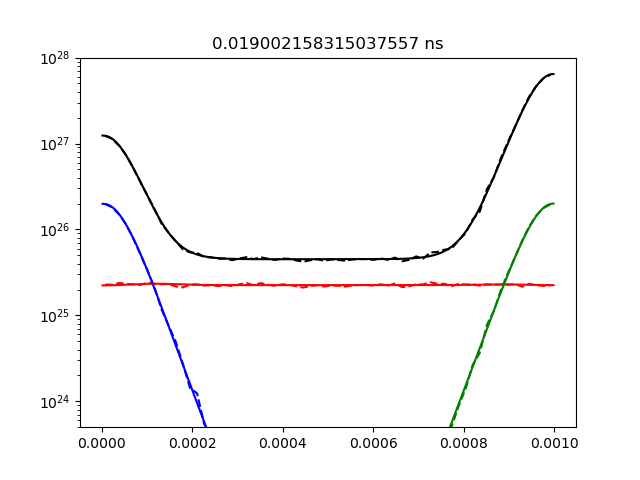}\includegraphics[width=0.33\columnwidth]{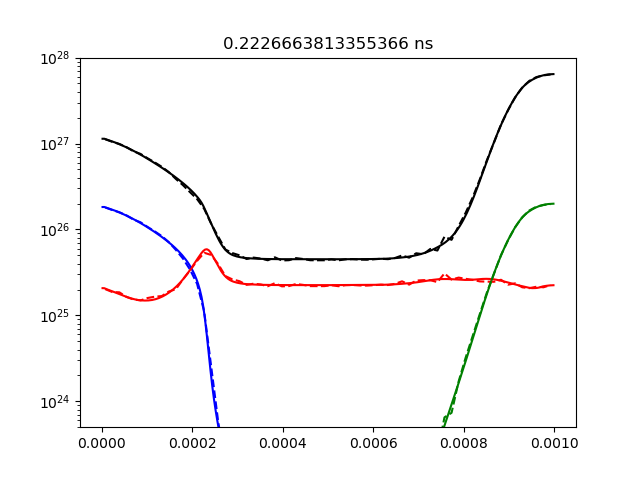}\includegraphics[width=0.33\columnwidth]{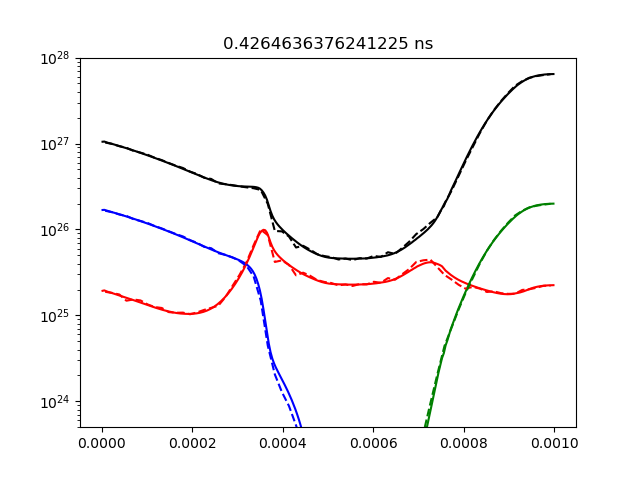}
\par\end{centering}
\begin{centering}
\includegraphics[width=0.33\columnwidth]{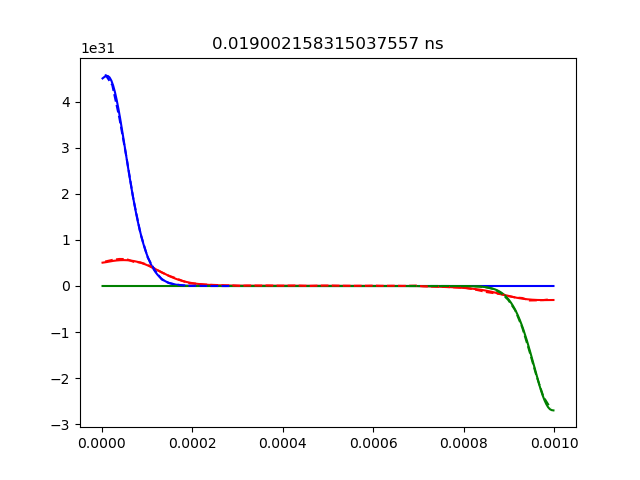}\includegraphics[width=0.33\columnwidth]{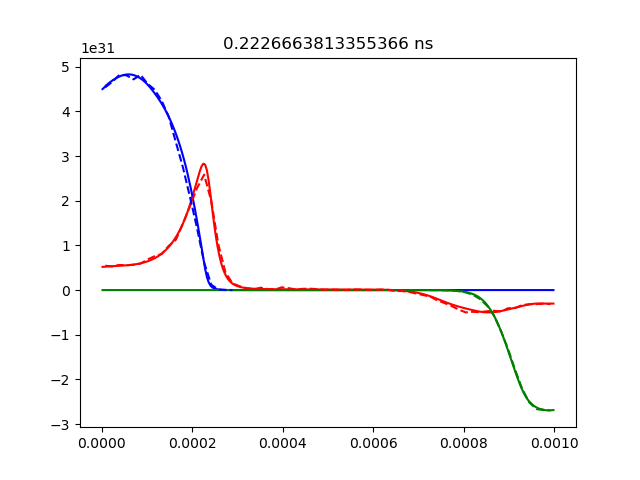}\includegraphics[width=0.33\columnwidth]{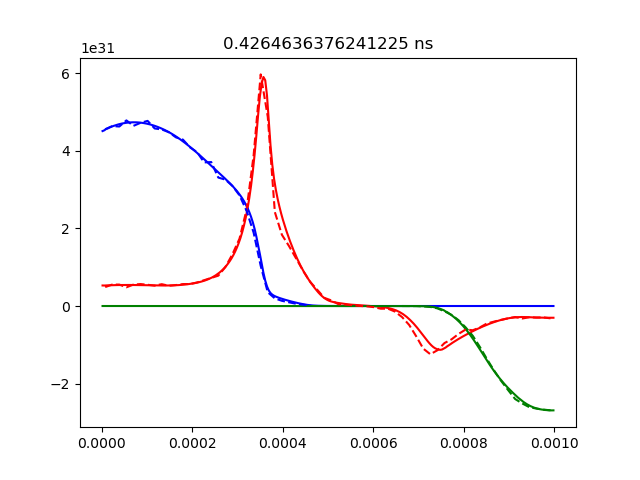}
\par\end{centering}
\begin{centering}
\includegraphics[width=0.33\columnwidth]{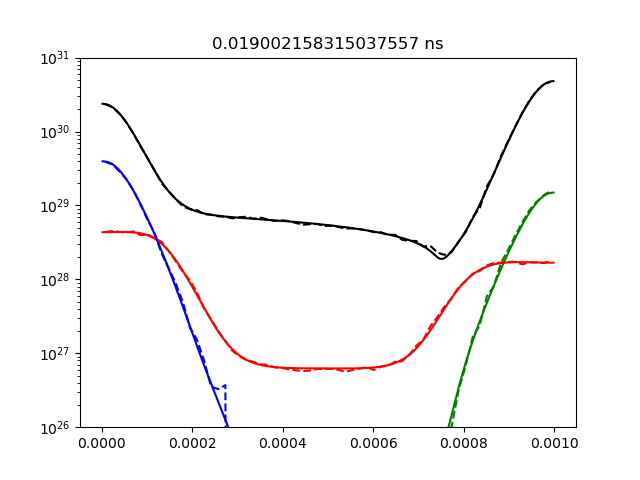}\includegraphics[width=0.33\columnwidth]{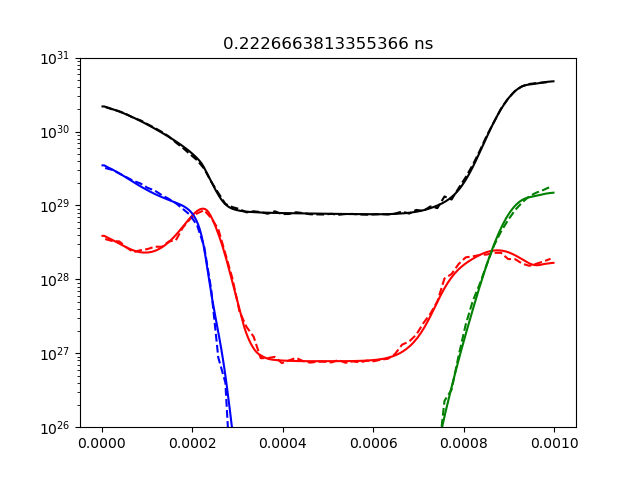}\includegraphics[width=0.33\columnwidth]{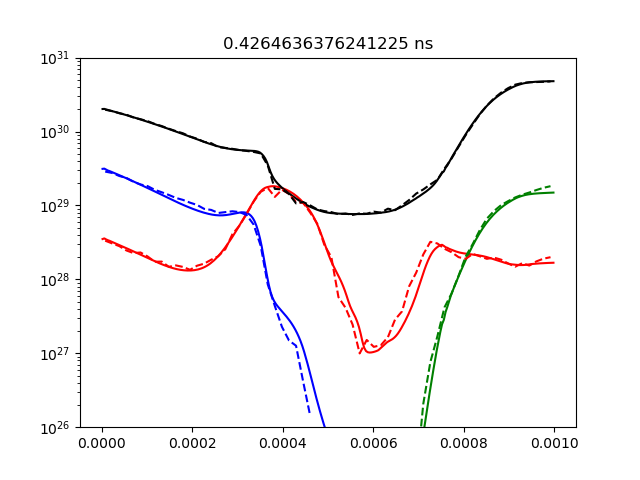}
\par\end{centering}
\caption{\protect\label{Fig:Lepape1d}1D simulation of the plasma interpenetration
problem. The solid lines depict results from iFP, while the dashed
lines depict the HMMC simulation. Moving from left to right, three
different time snapshots are captured. From top to bottom, the three
rows display density, momentum, and pressure. In the figure, black
is for electrons, green is for gold, blue is for carbon, and red is
for helium.}
\end{figure}

Figure \ref{Fig:Lepape1d} presents a comparison between the HMMC
simulation and iFP. We plot the first three moments (density, momentum,
and pressure) obtained from the simulations. Despite the different
discretizations and timestepping solvers in the two approaches, we
observe an overall excellent agreement between them throughout the
span of the simulation (0.42~ns, which is a very long time in these
types of experiments). The most precise agreement is observed in the
density, representing the zeroth moment. However, for higher moments,
the agreement remains strong at early times and gradually reveals
some minor differences, likely attributable to noise levels in low-density
regions arising from the limited number of particles available. Overall,
the level of agreement is remarkable between the two simulations,
lending credibility to the HMMC approach.\textcolor{red}{{} }The speedup
of HMMC vs. TA for this case is about \textcolor{red}{80}$\times$,
\textcolor{red}{with HMMC using $\nu_{Au}\Delta t=0.8$ and assuming
$\nu_{Au}\Delta t_{TA}=0.01$ for accuracy as before}.

\section{Discussion and summary}

\label{sec:conclusions}

\textcolor{black}{We have proposed a multiscale algorithm for multispecies
particle collisions, HMMC, that is able to improve performance vs.
classic MC methods by several orders of magnitude without loss of
long-term accuracy. HMMC considers Maxwellians for highly collisional
species and particles for moderately or weakly collisional ones, and
implements a multitiered collisional strategy comprising particle-particle
(with an extended TA algorithm), particle-Maxwellian (with an improved
Lemons algorithm), and Maxwellian-Maxwellian (with a five-moment model)
interactions. Each collisional interaction is strictly conservative,
and as a result the whole algorithm conserves mass, momentum, and
energy to numerical round-off. By considering a Maxwellian description
for highly collisional species, HMMC removes the stiff self-collision
timescale from the process, resulting in significant acceleration
vs. classical MC (TA).}

We have improved the individual component algorithms in various ways.
For TA MC particle-particle collisions, we have extended the algorithm
to use variable weights without loss of accuracy. For Lemons particle-Maxwellian
collisions, we have identified and fixed a failure mode in the limit
of small relative velocities between the particle and the Maxwellian
that enables accurate descriptions in arbitrary collisionality regimes
while employing larger timesteps. Finally, we have outlined how to
couple HMMC with a hybrid kinetic-ion/fluid-electron model, including
the form of the collisional sources needed for strict conservation,
and the form of the friction force to be applied to the ion species
from the interaction with the electron fluid.

We have demonstrated HMMC with challenging relaxation (0D, with both
fluid and kinetic species) and transport (1D, with a hybrid kinetic-ion/fluid-electron
description) problems, inspired from realistic conditions present
in ICF hohlraum environments (which typically feature many plasma
species of varying ionization levels, including highly ionized ones
that are particularly difficult to treat collisionally due to the
$Z^{4}$ scaling of the collision frequency). In all tests, we have
verified our implementation against the state-of-the-art hybrid Vlasov-Fokker-Planck
code iFP, finding excellent agreement. Our HMMC algorithm has been
shown to be at least two orders of magnitude faster for these applications
than classical MC without accuracy impact on the simulations, underscoring
its effectiveness for plasmas featuring varying collisionality regimes.

\section*{Acknowledgments}

This research has been funded by the Los Alamos National Laboratory
(LANL) Directed Research and Development (LDRD), Advanced Simulation
and Computation (ASC), and Inertial Confinement Fusion (ICF) programs.
The research used computing resources provided by the Los Alamos National
Laboratory Institutional Computing Program, and was performed under
the auspices of the National Nuclear Security Administration of the
U.S. Department of Energy at Los Alamos National Laboratory, managed
by Triad National Security, LLC under contract 89233218CNA000001.
The first author also receives partial support from the Exascale Computing
Project (17-SC-20-SC), a collaborative effort of the U.S. Department
of Energy Office of Science and the National Nuclear Security Administration.

\bibliographystyle{ieeetr}
\bibliography{hgmc}

\end{document}